\documentclass[]{article}
\usepackage{graphicx}
\usepackage{dcolumn}
\usepackage{bm}
\graphicspath{{Figures/}{./}}
\usepackage{lineno}
\usepackage{float}
\usepackage{booktabs} 
\usepackage{diagbox}
\usepackage[round]{natbib}
\usepackage{xcolor}
\usepackage{amsmath}
\usepackage{authblk}
\title{Testing the consistency of the resonant wave interaction approximation with simulated dynamics of idealized 2D internal wave fields}

\author[1]{Golan Bel}
\affil[1]{Department of Environmental Physics, Blaustein Institutes for Desert Research, Ben-Gurion University of the Negev, Sede Boqer Campus 8499000, Israel}
\author[2]{Eli Tziperman}
\affil[2]{Department of Earth and Planetary Sciences and School of Engineering and Applied Sciences, Harvard University, Cambridge, MA, USA}

\date{\today}
\graphicspath{{./}{./Figures/}}

\begin{document}

\maketitle

\begin{abstract}
Nonlinear interactions of internal ocean waves play an important role in ocean mixing and are of theoretical interest even in idealized settings. According to the resonant wave interaction approximation, which is commonly used to derive the kinetic equation for the energy spectrum, the dominant interactions are between wave triads whose wavevectors satisfy $\mathbf{k}=\mathbf{p}+\mathbf{q}$, and whose frequencies satisfy $\omega_{\mathbf{k}} = |\omega_{\mathbf{p}}\pm\omega_{\mathbf{q}}|$. In order to test the validity of the resonant wave interaction approximation, we examine several analytical derivations of the theory. The assumptions underlying each derivation are tested using idealized direct 2D, non-rotating numerical simulations, representing near-observed energy levels of the oceanic internal wave field. We show that the slow-amplitude assumptions underlying the derivations are inconsistent with the simulated dynamics in this particular set of simulations. In addition, most of the triads satisfying the resonant conditions do not contribute significantly to nonlinear wave energy transfer in our idealized 2D simulations, while some interactions that are dominant in nonlinear energy transfers do not satisfy the resonance conditions. Understanding the relevance of these results to the observed 3D rotating ocean internal wave field requires further study.
\end{abstract}


\section{Introduction}
\label{sec:introduction}

The resonant wave interaction approximation (RWIA) is used to simplify the study of nonlinearly interacting waves \cite[e.g.,][]{Phillips-1981:wave, Davis-Acrivos-1967:stability}. The approximation restricts interactions to wave triads (or quartets when triad interactions are not possible) that satisfy a condition on the wave frequencies of the form $\omega_{\mathbf{k}}=|\omega_{\mathbf{p}}\pm\omega_{\mathbf{q}}|$, in addition to the condition on the wavevectors $\mathbf{k}=\mathbf{p}+\mathbf{q}$. The wave triads that satisfy both conditions form ``resonance curves'' in the wavevector space, on which the significant interaction and energy transfer between waves with different wavevectors occur according to the theory. RWIA is used as part of the derivation of the kinetic equation of the weak wave turbulence formalism that aims to simplify the full nonlinear dynamics of weakly interacting wave fields \cite[][]{Hasselmann-1962:non, Zakharov-1965:weak, Zakharov-Lvov-Falkovich-2012:kolmogorov, Nazarenko-2011:wave}.

The RWIA-based weak wave turbulence formalism has been applied to surface ocean waves \cite[][]{Hasselmann-1962:non}, acoustic waves \cite[][]{Lvov-Lvov-Newell-et-al-1997:statistical}, internal waves in the ocean and atmosphere \cite[e.g.,][]{Mccomas-Bretherton-1977:resonant, Mueller-Holloway-Henyey-et-al-1986:nonlinear, Olbers-1976:nonlinear, Lvov-Polzin-Tabak-2004:energy}, and the study of Rossby-gravity wave interactions \cite[][]{Eden-Chouksey-Olbers-2019:mixed}. It has also been studied experimentally \cite[e.g.,][]{Herbert-Mordant-Falcon-2010:observation, Rodda-Savaro-Davis-et-al-2022:experimental, Rodda-Savaro-Bouillaut-et-al-2023:from}. The weak wave turbulence formalism is used to derive the closed kinetic equation for the energy spectrum. The resulting reduction in the number of interactions also leads to a more efficient computational treatment of the problem.

Our interest here is in the application of RWIA to the problem of a field of interacting internal waves in a stratified fluid. We critically examine the predictions of this approximation in the specific limited context of idealized 2D, non-rotating simulations. For internal ocean waves, the observed wave spectrum has been shown to have some universal properties represented by the Garrett-Munk (GM) empirical fit \cite[][]{Garrett-Munk-1972:space, Garrett-Munk-1975:space, Cairns-Williams-1976:internal, Garrett-Munk-1979:internal, Levine-2002:modification}, which was also studied in numerical simulations \cite[e.g.,][]{Pan-Arbic-Nelson-et-al-2020:numerical, Chen-Chen-Liu-et-al-2019:can} although deviations from the GM spectrum were observed \cite[][]{Wunsch-Webb-1979:climatology, Polzin-Lvov-2011:toward, Pollmann-2020:global, Le-Boyer-Alford-2021:variability}. This spectrum is not fully understood despite decades of research. Nonlinear internal wave interactions and the resulting energy transfers and wave breaking are responsible for much of the mixing in the ocean interior \cite[][]{Munk-Wunsch-1998:abyssal}. This mixing plays an important role in the absorption of CO$_2$ \cite[][]{Sabine-Feely-Gruber-et-al-2004:oceanic}, in forcing the large-scale meridional circulation, and consequently in the global meridional ocean heat transport \cite[][]{Bryan-1987:parameter, Nikurashin-Ferrari-2013:overturning}. The mixing occurs on scales smaller than can be represented in climate models. Its parameterization requires an understanding of the nonlinear wave interactions, which the RWIA and the accompanying weak wave turbulence formalism aspire to provide.

The key assumption underlying the RWIA is that the wave amplitudes vary over a timescale longer than the corresponding linear wave periods. Different derivations of the RWIA approach this condition differently and, therefore, result in different slowness criteria. To study the \textit{self-consistency} of the RWIA, one can initialize the RWIA-based kinetic equation for the internal wave field spectra with the GM spectrum and examine the timescale of the response. This (Boltzmann) timescale has been found to be inconsistent with the slowness assumption, posing a significant challenge to the theory \cite[][]{Holloway-1980:oceanic, Lvov-Polzin-Yokoyama-2012:resonant}. Taking into account near-resonance interactions \cite[][]{Lvov-Lvov-Newell-et-al-1997:statistical} was suggested to result in a slower and therefore self-consistent spectral adjustment timescale \cite[][]{Lvov-Polzin-Yokoyama-2012:resonant}. The ensemble-averaged rates of energy transfer due to wave-wave interactions in short, unforced, direct numerical simulations were found to be similar to those calculated using the kinetic equation \citep{Eden-Pollmann-Olbers-2019:numerical, Eden-Pollmann-Olbers-2020:towards}. The kinetic equation has also been shown to explain the transfer of energy between different regions of the spectrum \citep{Dematteis-Le-Pollmann-et-al-2024:interacting}. The RWIA has also been applied to study 2D internal waves, as we do here, and to derive the kinetic equation \citep{Ripa-1981:theory, Daubner-Zeitlin-1996:stationary, Caillol-Zeitlin-2000:kinetic, Shavit-Buehler-Shatah-2025:turbulent, Labarre-Shavit-2026:distinguished}. However, the question of the consistency of the kinetic equation and the assumptions used to derive it with the actual dynamics of typical oceanic internal wave fields, as opposed to the self-consistency of the kinetic equation, is still open.

In a recent work, \cite{Savva-Kafiabad-Vanneste-2021:inertia} developed a linear kinetic equation to describe the scattering of internal waves by an assumed steady quasi-geostrophic (QG) flow. They find that the scattering leads to a cascade to small scales along surfaces of constant frequency in wavenumber space. The cascade results in an internal wave spectrum of the form $k^{-2}$ for scales smaller than the scale of the internal wave forcing, a cascade that develops without nonlinear wave-wave transfers. This spectrum is similar to the spectra observed in the atmosphere and ocean, suggesting an alternative to the shaping of the spectrum by wave-wave interaction, which is our focus here. \cite{Savva-Kafiabad-Vanneste-2021:inertia} described the mechanism of interaction in the (WKBJ) limit, where internal wave wavelengths are smaller than QG scales, also analyzed by \cite{Kafiabad-Savva-Vanneste-2019:diffusion}, as a version of the induced diffusion mechanism of \cite{Mccomas-Bretherton-1977:resonant}, with the small-$K$ geostrophic mode playing the role of the low-frequency internal wave. They similarly identify the interaction of similar upward and downward propagating waves via scattering with the QG field as the elastic scattering mechanism of \cite{Mccomas-Bretherton-1977:resonant}. \citet{Barkan-Srinivasan-McWilliams-2024:eddy} analyzed the wave-eddy interaction using a coarse-graining approach in numerical simulations. They, again, show that the wave energy cascade is driven predominantly by scattering with eddies rather than via wave-wave interaction. \cite{Dong-Buehler-Smith-2020:frequency} used the WKBJ approximation, allowing for slow time variation of the QG flow, and found that internal wave energy diffuses both along and across constant frequency surfaces, and \citet{Dong-Buehler-Smith-2023:geostrophic} showed that even the interaction of waves with steady flow can lead to wave energy spreading off the constant frequency cone. However, \cite{Cox-Kafiabad-Vanneste-2023:inertia} examine the role of a time-dependent QG flow and find the scattered wave spectrum to still be localized within a thin boundary layer around the constant-frequency cone. While the scattering of internal waves by a stationary or slowly-varying QG flow may be a very efficient mechanism for cross-scale interactions, the fact that the resulting interactions are interpreted using terms that were derived based on the RWIA suggests that it is still worthwhile going back to this approximation and investigating its validity, as we do here. In addition, one does expect wave-wave interaction to dominate at small scales.

We follow three different derivations of the constraint on the frequencies of interacting waves represented by the RWIA that have been used as part of the weak wave turbulence formalism. Using direct numerical simulations (DNS) of idealized 2D internal wave fields, forced at low wavenumbers and run to a statistical steady state (Section~\ref{sec:details-numerical-simulations}), we test the specific assumptions in each derivation that lead to the above-mentioned constraint on the frequencies. We also use the DNS, performed with the flow\_solve model \cite[][]{Winters-MacKinnon-Mills-2004:spectral}, to examine all nonlinear interactions in wavenumber space. We find that the dominant interactions are not necessarily on the resonance curves. That is, the nonlinear interaction term calculated from the DNS shows that many of the resonant interactions do not contribute significantly to energy transfers, while there are important interactions that are non-resonant. We note that the idealized settings used do not allow making any claims regarding 3D or observed internal waves, and that the derivation of the statistical theory for nonlinear wave-wave interaction for oceanic internal waves is developed in 3D since its first appearance in \cite{Olbers-1976:nonlinear} as well as in more recent derivations \cite[e.g.,][]{Lvov-Tabak-2004:hamiltonian}.

The derivation of the kinetic equation typically relies on ensemble averaging. It is possible, in principle, to test it using DNS with different random initial conditions or different realizations of the stochastic forcing. However, our testing of the assumptions leading to the resonance condition relies on analyzing long, stochastically forced DNS that are at a statistical steady state. This implies that our testing involves time averaging rather than ensemble averaging.

\section{Details of the numerical simulations}
\label{sec:details-numerical-simulations}

Numerical simulations of the 2D Boussinesq equations (Appendix~\ref{sec:appendix-non-dimensionalization}) were performed with a double-periodic configuration of the flow\_solve pseudospectral model \cite[][]{Winters-MacKinnon-Mills-2004:spectral}. The size of the domain is $L_x=100$ km in the horizontal direction and $L_z=1$ km in the vertical direction. Most analyses use results with 512 grid points in the horizontal direction and 256 in the vertical direction, and we also show some results with double this resolution in both directions ($1024\times512$) to demonstrate robustness to resolution. A linear basic stratification is prescribed such that the Brunt–Väisälä buoyancy frequency is $N=(2\pi/30)$ minutes$^{-1}$, and the equations represent the full nonlinear dynamics of perturbations to this stratification (see Appendix~\ref{sec:appendix-non-dimensionalization}). We seek maximal simplification rather than relevance to ocean observations, and therefore set the Coriolis force to zero. Each integration was carried out to a statistical steady state, and then for 200 more days, which were used for the presented analyses. Our analyses are based on a high-frequency sampling of the model output at an interval of 18 s, allowing us to resolve all relevant motions and time derivatives from the output. The time step used for the integration is 1 s.

We use three runs with different amplitudes of temporally correlated stochastic forcing that is applied only to small wavenumbers of $0<|k_x|/(2\pi/L_x)\le2$ and $0<|k_z|/(2\pi/L_z)\le2$, where $\mathbf{k}=(k_x,k_z)$ is the wavevector. The forcing is applied to the buoyancy equation~\eqref{eq:buoyancy-equation-dim}, and no forcing was applied directly to the momentum equations~\eqref{eq:momentum-equations-dim}. The stochastic forcing for each wavevector was set using a temporally correlated red-noise with a $\tau=1$ day correlation time,
\begin{linenomath}
\begin{align*}
    R&=\exp(-\Delta t/\tau), \\
    \theta_n&=\theta_{n-1} R+\sigma \nu\sqrt{1-R^2},
\end{align*}
\end{linenomath}
where $\nu$ is a unit-amplitude random white noise and $\Delta t$ is the simulation time step. The stochastic forcing amplitudes of the different wavevectors are uncorrelated. The standard deviation of the medium forcing amplitude was three times larger than that of the weak one, and the forcing of the strong run was five times that of the weakly forced run.

The eddy viscosity and diffusivity were set to the minimal values, ensuring that a statistical steady state is achieved without numerical noise in the solution. The values used for the horizontal and vertical eddy viscosities are: $\tilde{\nu}_x=0.2$ m$^2$/s and $\tilde{\nu}_z=3\times10^{-5}$ m$^2$/s. The diffusivity was assumed to be isotropic and was set to $\tilde{\kappa}_x=\tilde{\kappa}_z=3\times10^{-5}$ m$^2$/s. We also simulated the dynamics using higher resolution of $1024\times512$, and for this simulation the viscosity and diffusivity are reduced to $\tilde{\nu}_x=2.5\times10^{-2}$, $\tilde{\nu}_z=1\times10^{-5}$, $\tilde{\kappa}_x=\tilde{\kappa}_z=1.0\times10^{-5}$ m$^2$/s.

\begin{figure*}
  \includegraphics[width=0.9\textwidth]{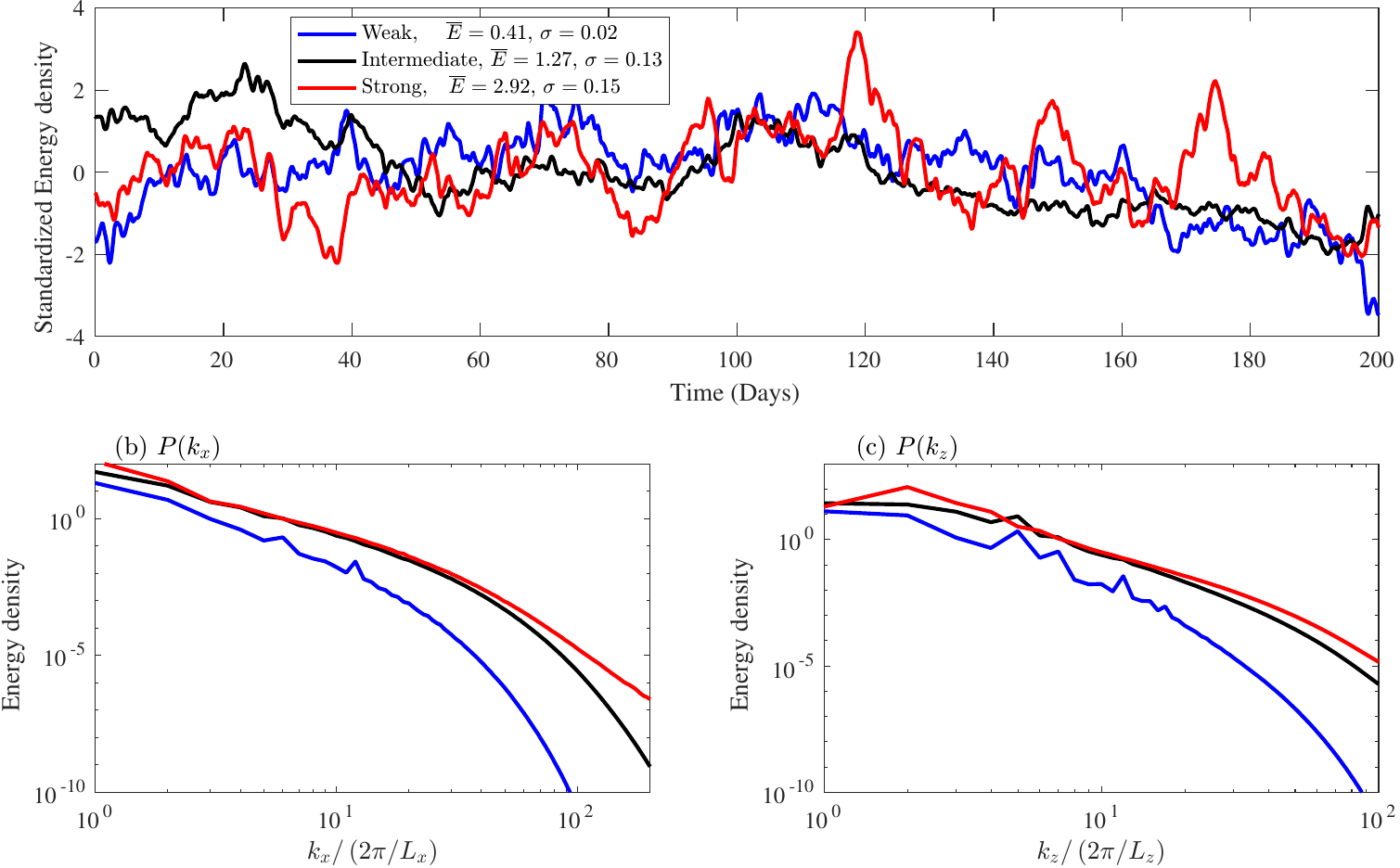}
  \caption{Characterizing the energy in the direct 2D numerical simulations: (a) Time series of standardized (mean removed, divided by std, with both specified in the figure legend) total energy per unit volume (J/m$^3$) for 200 days for the three runs. Energy spectra as a function of the horizontal wavenumber, $P(k_x)$, are depicted in panel (b), and energy spectra as a function of the vertical wavenumber, $P(k_z)$, are depicted in panel (c). The forced wave vectors are those with $0<|k_x|/(2\pi/L_x)\le2$ and $0<|k_z|/(2\pi/L_z)\le2$.}
    \label{fig:characterization}
\end{figure*}

Figure~\ref{fig:characterization} shows the time series of the total energy density and vertical and horizontal wavenumber energy spectra. The average energy levels of weakly, medium, and strongly forced model runs are 0.41, 1.27, and 2.92 J/m$^3$, respectively. The corresponding standard deviations (STDs) of the energy time series are 0.02, 0.13, and 0.15 J/m$^3$. Typical energy levels of ocean internal waves were estimated by \cite{Levine-2002:modification} as 0.85$N/N_\mathrm{ref}$ J/m$^3$, where $N_{\mathrm{ref}}=3$ cph. We use $N=2$ cph, and this implies an expected energy of 0.6 J/m$^3$, somewhere between the results of our simulations with weak and medium forcing amplitudes. Similarly, \cite{Lozovatsky-Morozov-Fernando-2003:spatial} showed an energy density of about 0.3\textendash2 J/m$^3$ depending on the distance from the mid-ocean bottom topography ridge where tidal energy is converted to internal waves. We therefore consider the energy levels of our weak and medium forcing cases to be in a range relevant to the dynamics of internal waves in the ocean. The simulated spectra in Fig.~\ref{fig:characterization}b show that the numerical eddy viscosity affects horizontal wavenumbers above 50 or so, where the spectra start decaying more rapidly beyond an inertial range not considerably affected by dissipation.

\section{Testing the results and assumptions behind derivations of the resonant-interaction approximation}

We start with an outline of the derivation of the energy equation written in terms of wave amplitudes (Section~\ref{sec:derivation-of-energy-equation}). Next, we proceed to our main result (Section~\ref{sec:finding}) which examines all interactions and shows that non-resonant interactions are important in our idealized 2D simulations, although these simulations are weakly nonlinear and one would expect them to be consistent with the RWIA. We also show that resonant interactions in these simulations are not necessarily important, although that is not a surprise, as it is a well-known result that special triads dominate among resonant interactions \cite[][]{Mccomas-Bretherton-1977:resonant}. Finally, in order to understand \textit{why} non-resonant interactions are important, we examine the assumptions behind three derivations of the approximation (Sections~\ref{sec:examining-first-derivation}, \ref{sec:examining-second-derivation}, \ref{sec:examining-third-derivation})

Different derivations of the RWIA rely on somewhat different assumptions, yet all are based on the wave amplitudes changing on a timescale longer than the linear wave period. Derivations of the RWIA have been presented using a continuous or discrete vertical representation of the internal wave field, using Eulerian, Lagrangian, and isopycnal coordinates, as well as a Hamiltonian formulation, and more 
\cite[][]{Hasselmann-1966:feynman, Kenyon-1968:wave, Mueller-Olbers-1975:dynamics, Mccomas-Bretherton-1977:resonant, McComas-1977:equilibrium, Pomphrey-Meiss-Watson-1980:description, Phillips-1981:wave, Milder-1982:hamiltonian, Caillol-Zeitlin-2000:kinetic, Lvov-Tabak-2001:hamiltonian, Lvov-Tabak-2004:hamiltonian}, see a helpful summary of approaches to deriving the kinetic equation of weak wave turbulence in Table~1 of \cite{Lvov-Polzin-Yokoyama-2012:resonant}. Regardless of the basic approach, all derivations then use a variant of the slowness assumption to derive the constraint on the frequencies of interacting waves. We discuss three different derivations of this constraint that are motivated by derivations in the literature and test the validity conditions for each, using our model simulations.

\subsection{Derivation of the energy equation in terms of wave amplitudes}
\label{sec:derivation-of-energy-equation}

We start by nondimensionalizing the momentum, buoyancy, and continuity equations (Appendix~\ref{sec:appendix-non-dimensionalization}) and derive (Appendix~\ref{sec:app-derivation-energy-equation}) nondimensional equations for the vorticity ($\zeta=u_{z}-w_{x}$) and buoyancy ($b=-g\rho/\rho_{0}$) in physical space, with $(x,z)$ and $(u,w)$ being the coordinates and velocities in the horizontal and vertical directions, respectively, and with $\rho$, $\rho_0$, and $g$ being the perturbation density relative to a specified linear mean stratification, a constant reference density, and gravity acceleration. Let the Fourier transform of the vorticity be $\zeta_{\mathbf{k}}$ and the Fourier transform of the buoyancy be $b_{\mathbf{k}}$ , where $\mathbf{k}=(k_x,k_z)$ is the wave vector. We define a wave amplitude in the spectral space, $A_{a,\mathbf{k}}$, \cite[e.g.,][]{Carnevale-1981:statistical} and calculate it from the output of our DNS,
\begin{linenomath}
\begin{align}
  A_{a,\mathbf{k}}=\frac{1}{2}(\zeta_{\mathbf{k}}/k+ab_{\mathbf{k}}),
\end{align}
\end{linenomath}
where $k$ is the magnitude of the wavevector and $a=\pm 1$, corresponding to right and left propagating waves. The dynamics of the wave amplitudes are described by, 
\begin{linenomath}
\begin{align}
   &\frac{\partial {A}_{a,\mathbf{k}}}{\partial t}
   +\mathsf{L}_{ab}A_{b,\mathbf{k}}  =
   \frac{\epsilon}{\left(4\pi\right)^{2}}
   {\int\limits_{-\infty}^{\infty}}
   d^{2}\mathbf{q}d^{2}\mathbf{p}\; \mathsf{M}_{abc}(\mathbf{k},\mathbf{p},\mathbf{q}) 
   \delta\left(  \mathbf{k-p-q}\right)
   A_{b,\mathbf{p}}A_{c,\mathbf{q}}+F_{a}(\mathbf{k}),
   \label{eq:A-equation}
\end{align}
\end{linenomath}
where $t$ denotes the time, and a sum over repeated indices ($b,c$) is implied; note that when $b$ is used as a subscript, it is an index taking the values $\pm1$ rather than referring to the buoyancy. Here, $\mathsf{L}=\mathsf{L}_w+\mathsf{L}_d$ is a linear operator representing the wave and dissipation parts, $\mathsf{M}_{abc}(\mathbf{k},\mathbf{p},\mathbf{q})$ is a nonlinear interaction coefficient (Appendix~\ref{sec:app-derivation-energy-equation}), and $\epsilon={u_{0}}/{NL_x}. $ is a dimensionless parameter characterizing the ratio of the nonlinear and linear parts of the amplitude dynamics (Appendix~\ref{sec:appendix-non-dimensionalization}). The exact energy equation, in terms of the wave amplitudes in spectral space and with no approximations applied yet, is obtained by multiplying \ref{eq:A-equation} by the complex conjugate of the wave amplitude, $A_{a,\mathbf{k}}^{\ast}$ (Appendix~\ref{sec:app-derivation-energy-equation}),
\begin{linenomath}
\begin{align}
   \label{eq:energy-equation} 
   &\frac{\partial}{\partial t}\left(  A_{a,\mathbf{k}}A_{a,\mathbf{k}}^{\ast}\right)
    =2\epsilon Re \Bigg(\frac{1}{4\left(2\pi\right)^{2}}
   {\int\limits_{-\infty}^{\infty}}
   d^{2}\mathbf{q}d^{2}\mathbf{p}\;
   {\color{blue}\delta\left(\mathbf{k}-\mathbf{p}-\mathbf{q}\right)
   \mathsf{M}_{abc}\left(\mathbf{k,p,q}\right)  A_{b,\mathbf{p}}A_{c,\mathbf{q}}A_{a,\mathbf{k}}^{\ast}}\Bigg) \nonumber \\
   &-2Re\left(L_{ab}A_{b,\mathbf{k}}A_{a,\mathbf{k}}^{\ast} \right)
    +2Re\left(F_{a}^{\ast}\left(  \mathbf{k}\right)  A_{a,\mathbf{k}}\right).
\end{align}
\end{linenomath}
The part of this equation highlighted in blue is the ``interaction term'' analyzed below, representing the transfer of energy between wavevectors $\mathbf{p},\mathbf{q}$ and $\mathbf{k}$. For linear dynamics and in the absence of dissipation ($\mathsf{L}_{d}=0$) and forcing ($F_a=0$), the energy of any wave mode, right or left propagating ($a=\pm1$), is conserved.

\subsection{Finding: some non-resonant interactions are important}
\label{sec:finding}

The resonance condition implies that the nonlinear interaction term on the RHS of the energy equation, $\mathsf{M}_{abc}\left(\mathbf{k,p,k-p}\right) A_{b,\mathbf{p}}A_{c,\mathbf{k-p}}A_{a,\mathbf{k}}^{\ast}$, is strong only if 
\begin{align}
   \Delta\omega(\mathbf{k},\mathbf{p},\mathbf{q},a,b,c)
   =a\omega_{\mathbf{k}}-b\omega_{\mathbf{p}}-c\omega_{\mathbf{q}}=0,
   \label{eq:delta_omega=0}
\end{align}
where $a,b,c=\pm1$ indicate the wave phase propagation direction, in addition to the condition on the wavevectors imposed by the delta function $\delta\left(\mathbf{k-p-q}\right)$. 

\begin{figure*}
\includegraphics[width=\textwidth]{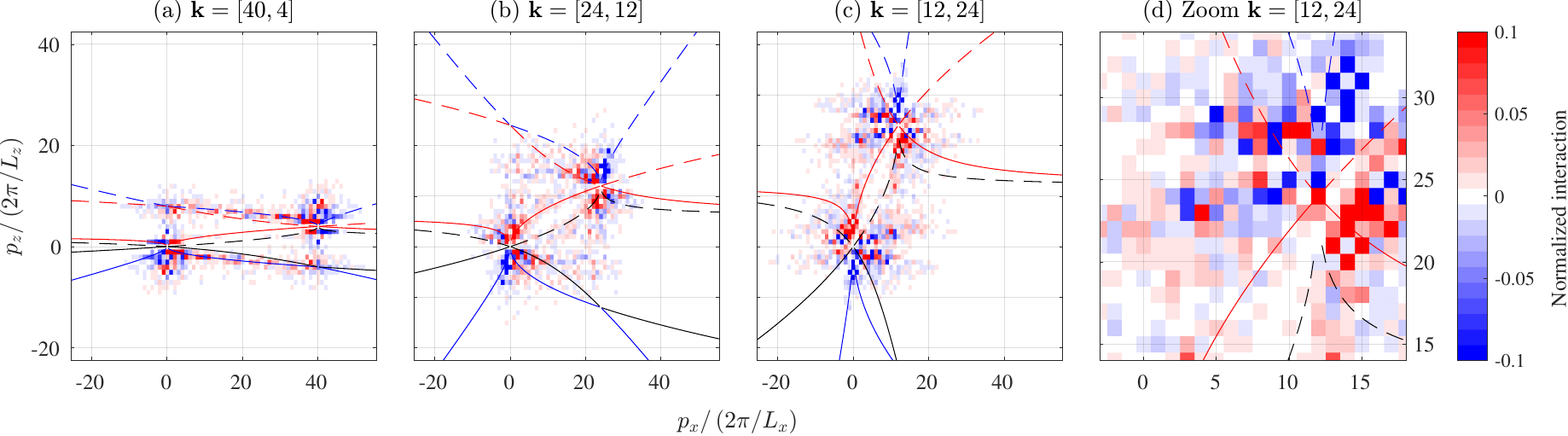}
\caption{(a--c) Time-average of the interaction term ($\mathsf{M}_{abc}\left(\mathbf{k,p,k-p}\right) A_{b,\mathbf{p}}A_{c,\mathbf{k-p}}A_{a,\mathbf{k}}^{\ast}$ in eq.~\eqref{eq:energy-equation}) for nondimensional wavevectors $\mathbf{k}=(40,4)$, $(24,12)$, and $(12,24)$, for the medium forcing amplitude run, normalized (for each $\mathbf{k}$) by its maximum value. The color range corresponds to a part of the range of values plotted to include the weakest 10\% of the interactions, which account for about 50\% of the total energy transfers, as shown by Fig.~\ref{fig:interaction-CDF}. The positive values (red) correspond to gain and the negative ones (blue) to loss of energy by wavevector $\mathbf{k}$. (d) A zoom into the region of $\mathbf{p}\sim\mathbf{k}$ for panel (c). The dashed and solid, blue, red, and black lines represent the different resonance curves, see the text for details. The forced wave vectors are those with $0<|k_x|/(2\pi/L_x)\le2$ and $0<|k_z|/(2\pi/L_z)\le2$.}
\label{fig:interaction-terms_27_climn=0.1}
\end{figure*}

Fig.~\ref{fig:interaction-terms_27_climn=0.1} shows the time average of the nonlinear interaction term\\ $\mathsf{M}_{abc}\left(\mathbf{k,p,k-p}\right) A_{b,\mathbf{p}}A_{c,\mathbf{k-p}}A_{a,\mathbf{k}}^{\ast}$ in eq.~\eqref{eq:energy-equation} for the simulation with medium-amplitude forcing, for three different wavevectors $\mathbf{k}$. The lines are solutions to $\mathbf{k}=\mathbf{p}+\mathbf{q}$ and the resonance condition $a\omega_{\mathbf{k}}-b\omega_{\mathbf{p}}-c\omega_{\mathbf{q}}=0$, which may be reduced to $-\hat{\omega}_\mathbf{k}+\hat{\omega}_\mathbf{p}+\hat{\omega}_\mathbf{q}=0$, $\hat{\omega}_\mathbf{k}+\hat{\omega}_\mathbf{p}-\hat{\omega}_\mathbf{q}=0$ or $\hat{\omega}_\mathbf{k}-\hat{\omega}_\mathbf{p}+\hat{\omega}_\mathbf{q}=0$ where the hatted frequencies are positive, $\hat{\omega}_\mathbf{k}=|k_x|/\sqrt{k_x^2+k_z^2}$. These are quadratic equations for, say $p_z$ as a function of $p_x$ ($\mathbf{k}$ is specified), and, therefore, there are six solutions for the resonant $p_z$, corresponding to the six line types in the figure. The resonance corresponding to $\hat{\omega}_\mathbf{q}=\hat{\omega}_\mathbf{k}-\hat{\omega}_\mathbf{p}$ is shown by the blue curves, the one satisfying $\hat{\omega}_\mathbf{q}=\hat{\omega}_\mathbf{k}+\hat{\omega}_\mathbf{p} $ in red, and   $\hat{\omega}_\mathbf{q}=\hat{\omega}_\mathbf{p}-\hat{\omega}_\mathbf{k}$ in black. Each resonance has two solutions, shown by the solid and dashed curves. With $\mathbf{k}$ specified, the interaction term becomes a function of $\mathbf{p}=(p_x,p_z)$ alone, and its average is shown by the color shading; as can be easily seen, it is of a large magnitude only for a small fraction of these resonant wavevectors. This interaction term, based on the triple product of wave amplitudes, is predicted by the weak wave turbulence theory to be dominated by its values on the resonant curves, as predicted, for example, by eq.~(2.1.11) in \cite{Zakharov-Lvov-Falkovich-2012:kolmogorov}, at the limit of vanishing friction.

\begin{figure*}
\includegraphics[width=\textwidth]{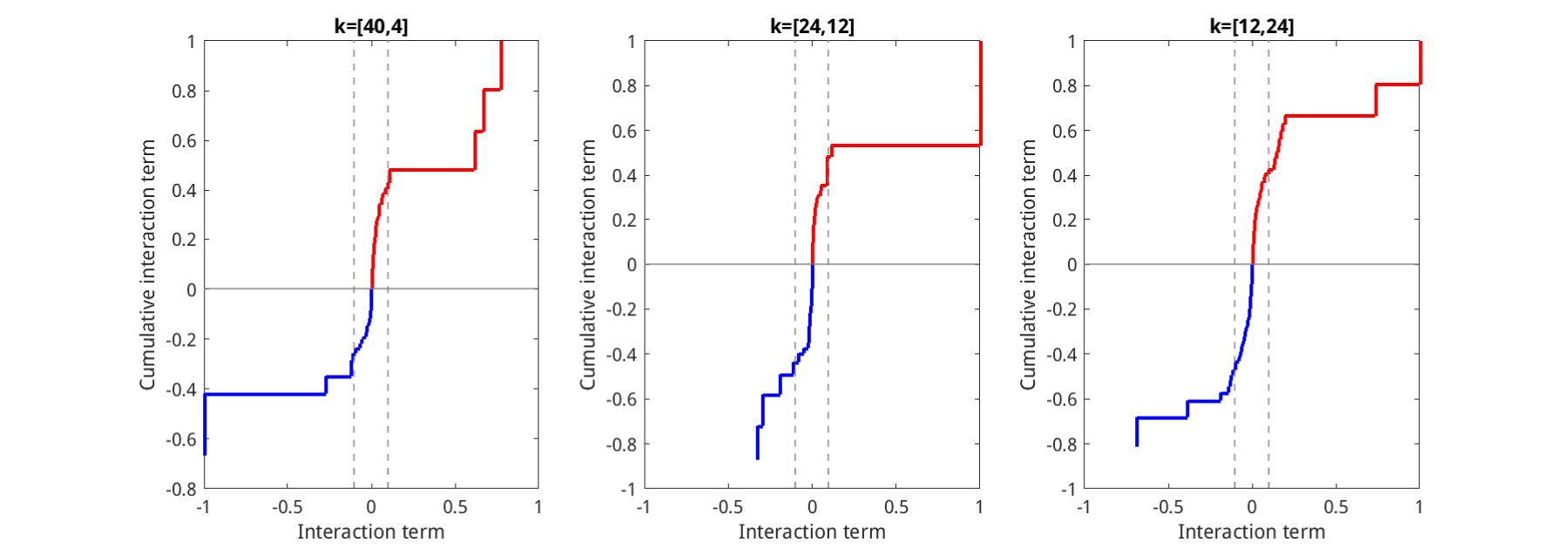}
\caption{The cumulative sum over the interaction term for the weakly forced experiment that is shown in Fig.~\ref{fig:interaction-terms_28_climn=0.1}, which is the weak forcing equivalent of Fig.~\ref{fig:interaction-terms_27_climn=0.1}. The cumulative sum is calculated for positive interaction values as the cumulative sum over the sorted interaction values, and for negative values similarly as the cumulative sum over the sorted negative values. The cumulative sum is then normalized by the largest in absolute value of the positive and negative cumulative sums. The rate of change of the energy of the wavenumber shown in the title of each panel is the sum over all the interaction values shown in Fig.~\ref{fig:interaction-terms_28_climn=0.1}. Therefore, the cumulative sum shows that about 40--60\% of this rate of change is achieved by the sum over interaction values smaller than 10\% of the maximum interaction term values, bounded by the vertical dash lines.}
\label{fig:interaction-CDF}
\end{figure*}

We find that some non-resonant interactions are stronger than most resonant interactions. These are well-demonstrated for the medium-amplitude forcing run in the zoomed-in Fig.~\ref{fig:interaction-terms_27_climn=0.1}d, and in Figs.~\ref{fig:interaction-terms_28_climn=0.1} and \ref{fig:interaction-terms_29_climn=0.1} which show the average of the interaction term for the weakly and strongly-forced simulations. One can see darkly colored pixels away from the resonance curves, involving unforced wavevectors $\mathbf{k}, \mathbf{p}$, and $\mathbf{q}$. The color bar range highlights weaker interactions: the interaction term is normalized by its maximum absolute value (the maximum is with respect to the different $p_x,p_z$ values), and the color range corresponds to 10\% of this maximum value, allowing to see even the weaker interactions. Fig.~\ref{fig:interaction-CDF} shows the cumulative sum of the interaction term, showing that these 10\% weakest interactions correspond to about 50\% of the total energy transfers, justifying this color range choice. Corresponding supplementary Figures~\ref{fig:interaction-terms_28_climn=1}, \ref{fig:interaction-terms_27_climn=1}, and \ref{fig:interaction-terms_29_climn=1} use color scales spanning the full range of the interaction term.

\begin{figure*}
\includegraphics[width=\textwidth]{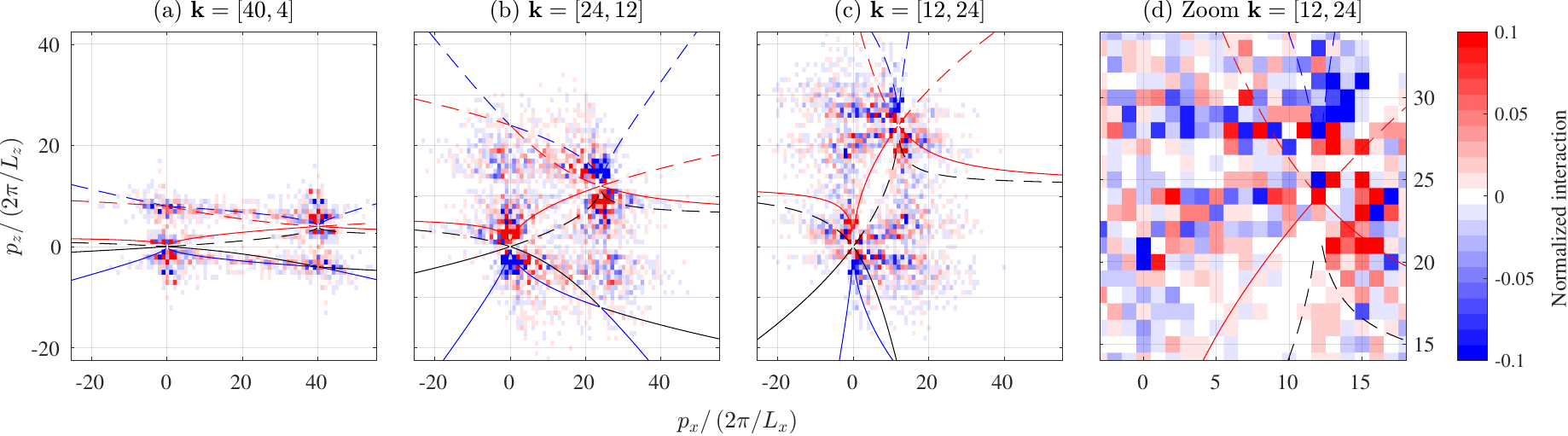}
\caption{(a--c) Time-average of the interaction term (the expression in blue on the second line of eq.~\eqref{eq:energy-equation}) 
for $\mathbf{k}=(40,4),~(24,12)$, and $(12,24)$, for the \textbf{doubled resolution} ($1024\times512$) run. The mean energy of this run is 0.7 J/m$^3$, which is between the weak- and medium-forced runs shown in Fig.~\ref{fig:characterization}. The shown interaction values are normalized (for each $\mathbf{k}$) by their maximum, and the color range shows the weakest 10\% to see a wider range of interacting wavenumbers as discussed in Fig.~\ref{fig:interaction-terms_27_climn=0.1}. (d) A zoom into the region of $\mathbf{p}\sim\mathbf{k}$ for panel (c). The lines represent the resonance curves.}
\label{fig:interaction-terms_1_climn=0.1}
\end{figure*}

The values of the nondimensional nonlinearity parameter $\epsilon$ (Appendix~\ref{sec:appendix-non-dimensionalization}) for the three simulations are small: $6\cdot10^{-5}$, 0.0001, 0.0002, for the weak, intermediate, and strong forcing, respectively, suggesting that the RWIA should apply to all three runs, while the above interaction term plots suggest that this is not the case. Fig.~\ref{fig:interaction-terms_1_climn=0.1} shows the average of the interaction term for a model run at twice the spatial resolution. As in the above figures, many non-resonant pixels are highlighted, suggesting that the issues with RWIA in our idealized 2D simulations are robust to the resolution. As a reminder, only wavevectors with nondimensional wave numbers $0<|k_x|/(2\pi/L_x)\le2$ and $0<|k_z|/(2\pi/L_z)\le2$ are forced (Section~\ref{sec:details-numerical-simulations}). That is, only the wavevectors very close to the origin in Fig.~\ref{fig:interaction-terms_27_climn=0.1} are directly forced. In order to test the sensitivity to the DNS resolution, we show in Fig.~\ref{fig:interaction-terms_1_climn=0.1} the averaged interaction term for a case with double the resolution ($1024\times512$, see details in Section~\ref{sec:details-numerical-simulations}). The results are robust: non-resonant interactions are still significant.

Interestingly, strong interactions (darker blue or red pixels) occur for triads in which two wavevectors are large and one is small (i.e., for $\mathbf{p}\approx\mathbf{k}$, or, equivalently, $\mathbf{q}\approx\mathbf{k}$). The parametric subharmonic instability and induced diffusion interactions \cite[][]{Mccomas-Bretherton-1977:resonant} also involve a small wavevector mode, although the strong interactions in the above figures do not necessarily lie on the resonant lines where these special triads are defined. Furthermore, these special triad interactions are relevant only when the RWIA is valid, while we show indications below that the required conditions for the RWIA to hold are not met in our simulations. Another possible explanation of some of the dominant interactions in Fig.~\ref{fig:interaction-terms_27_climn=0.1}, between \textit{forced} small wavevectors and larger ones, may be the generalized quasi-linear approximation \citep[GQL,][]{Marston-Chini-Tobias-2016:generalized}. However, the same Fig.~\ref{fig:interaction-terms_27_climn=0.1}, (as well as the double resolution case in Fig.~\ref{fig:interaction-terms_1_climn=0.1} and strongly forced run shown in Fig.~\ref{fig:interaction-terms_29_climn=0.1}) also shows strong interactions even between unforced wavevectors. It is, therefore, not clear that GQL approximation describes the dominant interactions for our three simulations.

We now consider three different derivations of the RWIA starting from the above equations and test their assumptions against our simulations. Our goal is to understand why the RWIA does not explain the entire set of important interactions in our idealized 2D simulations.

\subsection{Examining a first derivation of the resonant interactions approximation}
\label{sec:examining-first-derivation}

The \textbf{first} derivation considered here is based on a formal two-timescale perturbation approach applied to the amplitude equation \eqref{eq:A-equation}. For details on this method, see \citet[][]{Benney-Saffman-1966:nonlinear}, and Section 11.2 of \citet[][]{Bender-Orszag-1978:advanced}, or, in the context of the weak wave turbulence approach to internal waves in, e.g., \citet[][]{Caillol-Zeitlin-2000:kinetic}. We start by assuming that the solution can be expanded in the small parameter, $\epsilon$, namely,
\begin{linenomath}
\begin{align}
   A_{a,\mathbf{k}}=A_{a,\mathbf{k,}0}
   +\epsilon A_{a,\mathbf{k,}1}+\epsilon
   ^{2}A_{a,\mathbf{k,}2}+\cdots,
\end{align}
\end{linenomath}
and the slow time scale is defined as $\tau=\epsilon t$. 
The zeroth order variable simply satisfies the linear equation,
\begin{linenomath}
\begin{align}
   \frac{\partial}{\partial t}
   A_{a,\mathbf{k,}0}+L_{ab}A_{b,\mathbf{k,}0}
   =F_{a}\left(\mathbf{k}\right),
\end{align}
\end{linenomath}
with a solution for $\mathbf{k}$ away from the forced wavevectors of the form (neglecting dissipation and diffusion),
\begin{linenomath}
\begin{align}
  A_{a,\mathbf{k,}0} =\hat{A}_{a,\mathbf{k}}\left(  \tau\right)
  e^{-ia\omega_\mathbf{k}t}.
  \label{eq:slow-amplitude-definition}
\end{align}
\end{linenomath}
The equation for the first order is
\begin{linenomath}
\begin{align}
   &\frac{\partial}{\partial t}
   A_{a,\mathbf{k,}1}+L_{ab}A_{b,\mathbf{k,}1}  
   =-\frac{\partial}{\partial\tau}
   A_{a,\mathbf{k,}0} \nonumber \\
   & +\frac{1}{4\left(2\pi\right)^{2}}
   {\displaystyle\int\limits_{-\infty}^{\infty}}
   d^{2}\mathbf{q}d^{2}\mathbf{p}
   \delta\left(  \mathbf{k-p-q}\right)
   \mathsf{M}_{abc}\left(\mathbf{k,p,q}\right)  A_{b,\mathbf{p,}0}A_{c,\mathbf{q,}0}.
\end{align}
\end{linenomath}
The RHS can be shown (Appendix~\ref{sec:app-2-time-scale}) to be proportional to $e^{-ia\omega(\mathbf{k})t}$, which is the solution to the LHS operator and, thus, may result in a secular term that grows linearly in $t$, a growth we want to bound so that the order $\epsilon$ terms remain smaller than the zeroth order and the perturbation series remains valid for long times. Requiring the coefficient of the secular term to vanish leads to the resonant interaction condition represented by eq.~\eqref{eq:delta_omega=0}, consistent with the RWIA whose testing is our focus here.

Consider a couple of caveats. For continuous fields of random waves, the bounded growth condition is applied to the \textit{energy density} \citep[see the discussion paragraph at the end of section 2.4 of][]{Benney-Saffman-1966:nonlinear}. For discrete spectra, such as those appearing in our finite domain simulations, the secular terms can be eliminated in the equation for the \textit{amplitudes} for the perturbation expansion to be valid for large $t$ (see the above mentioned discussion paragraph). There are, of course, other ways of defining a slowly-varying wave amplitude, including a running-mean of ${A}_{a,\mathbf{k}}$, or the absolute value of $\hat{A}_{a,\mathbf{k}}$. However, the derivations considered in this paper, including the one considered here, use the definition of eq.~\eqref{eq:slow-amplitude-definition}. Our objective is to test these derivations, and therefore we use this definition.

\begin{figure*}
\includegraphics[width=\textwidth]{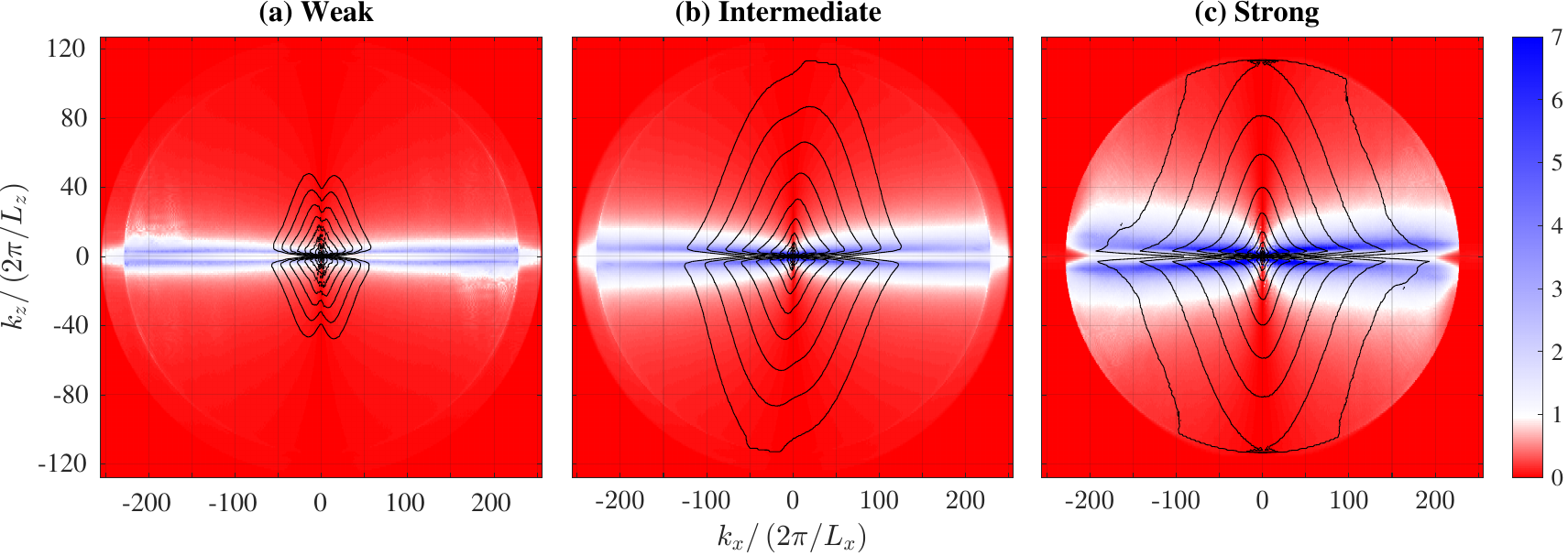}
\caption{Testing the slow amplitude assumption, requiring $V_1\gg1$, where $V_1$ is the ratio of the time scale of the slow amplitude and the linear wave period (eq.~\ref{eq:V1}). Color shading shows $V_1$ and the black contours show the spectral energy $\log_{10}E(\mathbf{k})$ with contour lines drawn from $-8$ to 2 with a contour interval of 1. (a) weak, (b) intermediate, and (c) strong forcing. The contour lines show that there is significant energy in wavenumbers that do \textit{not} satisfy the slowness condition (red shading), spanning most resolved wavenumbers.}
\label{fig:V1}
\end{figure*}

This derivation depended on the wave amplitude $\hat A_{a,\mathbf{k}}$ varying with $\tau=\epsilon t$, and thus on a longer time scale than the linear wave period. To test whether such a slow amplitude approximation is consistent with our numerical simulations, we define a measure $V_1(\mathbf{k},a)$, the ratio between the timescale of slow amplitude variations and the linear wave period. The time scale of the slow amplitude variations, $T(\hat{A}_{a,\mathbf{k}})$, is estimated via the RMS of its time derivative. The self-consistency of the first derivation of the RWIA requires that this measure satisfies,
\begin{linenomath}
\begin{align}
    V_1(\mathbf{k},a)
    =\frac{T(\hat{A}_{a,\mathbf{k}})}{{2\pi}/|\omega_{\mathbf{k}}|}
    =\left(\frac{1}{[\hat{A}_{a,\mathbf{k}}]} 
    \left[\frac{d\hat{A}_{a,\mathbf{k}}}{dt}\right]\right)^{-1}
     {|\omega_{\mathbf{k}}|}\gg1.
    \label{eq:V1}
\end{align}
\end{linenomath}
Here $[\cdot]$ indicates the RMS of the real part of the quantity in the square brackets (the results for the imaginary part, which are not presented here, are similar). Fig.~\ref{fig:V1} shows $V_1$ for all the resolved wavevectors, and clearly, for nearly all wavevectors except those with a small vertical wavenumber, it is not larger than one (the range smaller than one is shown by red shading), certainly not \textit{much} larger. The contours in Fig.~\ref{fig:V1} show the spectral energy density $\log_{10}E(\mathbf{k})$, making it clear that wavevectors with significant energy do not satisfy the slow-amplitude condition. 

\begin{figure*}
\centering
\includegraphics[width=\textwidth]{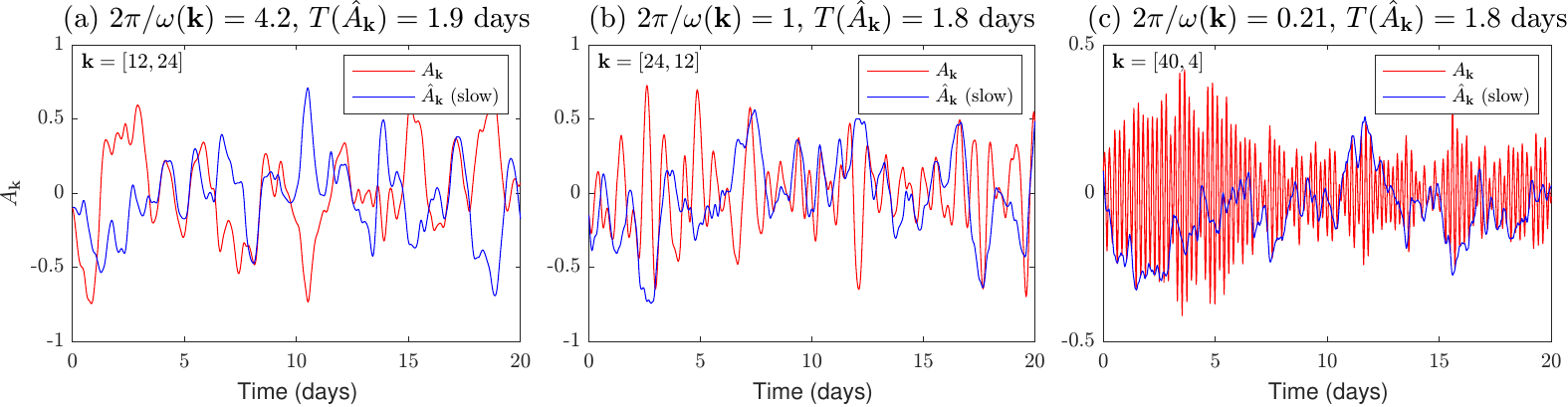}
\caption{Time series of the fast and slow amplitudes of three particular wavevectors in the weakly forced simulation. These panels demonstrate that the slow amplitude (blue) varies on a time scale not much longer than that of the linear wave period, as also quantified in Fig.~\ref{fig:V1}.}
\label{fig:Fig_Ap_Ap_slow_timeseries_expnum_28}
\end{figure*}

The reason for this is made clear in Fig.~\ref{fig:Fig_Ap_Ap_slow_timeseries_expnum_28}. The figure shows the fast and slow amplitudes for the weakly forced run for three different wavevectors. Even for the relatively short period waves in panel (c), the slow amplitude envelope shown by the blue line is not smooth, and its time scale is therefore not larger than the time scale of the oscillating part. For the longer-period waves in the other two panels, it is difficult to distinguish between the wave amplitude and what was supposed to be the slow amplitude. A similar picture is seen for the more strongly forced experiments (Figs.~\ref{fig:Fig_Ap_Ap_slow_timeseries_expnum_27}, \ref{fig:Fig_Ap_Ap_slow_timeseries_expnum_29}). In addition, Supplementary Table~\ref{tab:time-scale-A-and-abs_A} shows the time scales calculated from the simulation for the same three wave numbers analyzed in Fig.~\ref{fig:interaction-terms_28_climn=0.1} for the real, imaginary, and absolute value of the slow wave amplitude. The results again indicate that these different time scale measures of the slow amplitude are not consistently much larger than the time scale of the linear wave.

\subsection{Examining a second derivation of the resonant interactions approximation}
\label{sec:examining-second-derivation}

The standard derivation of the RWIA-based kinetic equation, in the context of weak wave turbulence, is based on the energy rather than the amplitude equation and typically uses ensemble averaging and a Gaussian decomposition of a fourth-order amplitude product. The \textbf{second} derivation of the RWIA we consider is based on the energy equation, while still avoiding ensemble averaging and the Gaussian assumption. We proceed by separating the wave amplitudes into a slow part and an oscillating term, ${A}_{a,\mathbf{k}}(t)=\hat{A}_{a,\mathbf{k}}(t)e^{-ia\omega_{\mathbf{k}}t}$, temporally averaging the energy equation \eqref{eq:energy-equation} over a time $T$, and neglecting dissipation and forcing terms, to find (Appendix~\ref{sec:app-slow-tripple-correlation}),
\begin{linenomath}
\begin{align}
    \label{eq:TA_denergy_dt}
    &\frac{1}{T}\left.\left(  \hat{A}_{a,\mathbf{k}}\hat{A}_{a,\mathbf{k}}^{\ast}\right)\right|_0^T
    = Re\Bigg(\frac{2\epsilon}{\left(4\pi\right)^{2}}
   {\int\limits_{-\infty}^{\infty}}
   d^{2}\mathbf{q}d^{2}\mathbf{p}\nonumber \\
   &
   \times\delta\left(\mathbf{k-p-q}\right)
   {\sum\limits_{b,c}}
   \mathsf{M}_{abc}\left(  \mathbf{k,p,q}\right)
   \times\frac1T\int\limits_0^T
   \hat{A}_{b,\mathbf{p}}\hat{A}_{c,\mathbf{q}}\hat{A}_{a,\mathbf{k}}^{\ast}
   e^{-i\left(-a\omega_{\mathbf{k}}+b\omega_{\mathbf{p}}+c\omega_{\mathbf{q}}\right)t}dt
   \Bigg).
\end{align}
\end{linenomath}
Now assume that the product of the three slow amplitudes varies slower than the oscillating exponent in the last line of this equation. The slowness of the amplitudes is used to simplify the integral in the last line, which we denote as:
\begin{align*}
    Y_{a,b,c}(\mathbf{k},\mathbf{p},\mathbf{q})=Re\left(\frac1T \int_0^T\hat{A}_{b,\mathbf{p}}\hat{A}_{c,\mathbf{q}} \hat{A}_{a,\mathbf{k}}^{\ast} e^{-i\left(-a\omega_{\mathbf{k}} +b\omega_{\mathbf{p}} +c\omega_{\mathbf{q}}\right)t}dt\right).
\end{align*}
Under this slowness assumption, we can separate the slow triple product and the oscillating term and write the averaged expression as the product of averages. In the limit $T\to\infty$, the average over the exponential term leads to the term $\delta_{0,\Delta\omega}$ below, which is 1 if $\Delta\omega=0$ and zero otherwise, thus enforcing the resonance condition given by eq.~(\ref{eq:delta_omega=0}). Denoting the time average over a time $T$ as $\overline{(\cdot)}$, this approximation can be written as:
\begin{linenomath}
\begin{align}
   \label{eq:Slow_Amp_Approx_3prod}
   X_{a,b,c}(\mathbf{k},\mathbf{p},\mathbf{q})&\equiv Re\Bigg(\overline{\hat{A}_{b,\mathbf{p}} \hat{A}_{c,\mathbf{q}} \hat{A}_{a,\mathbf{k}}^{\ast}}
   \;\times\;
  \frac{1}{T}\int\limits_0^T
   e^{i\Delta\omega(\mathbf{k},\mathbf{p},\mathbf{q},a,b,c) t}dt\Bigg)\nonumber \\
       &\underset{T\to\infty}{\to}
      Re\left(\overline{\hat{A}_{b,\mathbf{p}} \hat{A}_{c,\mathbf{q}} \hat{A}_{a,\mathbf{k}}^{\ast}}
   \right) 
   \delta_{0,\Delta\omega},
\end{align}
\end{linenomath}
where $\Delta\omega(\mathbf{k},\mathbf{p},\mathbf{q},a,b,c)$ is defined in eq.~(\ref{eq:delta_omega=0}). The two quantities, $X$ and $Y$, should be approximately the same for the RWIA derivation to be valid. We now define a quantitative measure of the difference between these two quantities. Crudely, this measure can be thought of as being the difference between $X$ and $Y$ divided by the maximum of the two, and calculated for each wavevector $\mathbf{p}=(p_x,p_z)$.

Fixing $\mathbf{k}$, and given the delta function over the three wavevectors which imposes $\mathbf{q}=\mathbf{k}-\mathbf{p}$, the difference between $X$ and $Y$ becomes a function of only the components of $\mathbf{p}=(p_x,p_z)$ and the three indices $a,b,c$. To quantify the difference between the two, we use the following measure, which, for each value of $(p_x,p_z)$ looks at the maximum over all values of $a,b,c$ of the absolute value of the difference between $X$ and $Y$.
\begin{linenomath}
\begin{align}
  V_{2}(\mathbf{k},\mathbf{p},\mathbf{q})
  &={\rm sign}\left(X_{a*,b*,c*}(\mathbf{k},\mathbf{p},\mathbf{q})-Y_{a*,b*,c*}(\mathbf{k},\mathbf{p},\mathbf{q}) \right) \nonumber \\
  &\times \max_{a,b,c}\frac{|X_{a,b,c}(\mathbf{k},\mathbf{p},\mathbf{q})-Y_{a,b,c}(\mathbf{k},\mathbf{p},\mathbf{q})|}
  {\mathrm{normalization}(\mathbf{k},\mathbf{p})}
  \label{eq:V2}
\end{align}
\end{linenomath}
The normalization factor in the denominator is calculated as the maximum of $|X|$ and $|Y|$ over all $a,b,c$ values,
\begin{linenomath}
\begin{align*}
    \mathrm{normalization}(\mathbf{k},\mathbf{p})
    ={\displaystyle\max_{a,b,c}\left(|X_{a,b,c}(\mathbf{k},\mathbf{p},\mathbf{q})|,|Y_{a,b,c}(\mathbf{k},\mathbf{p},\mathbf{q})|\right)}.
\end{align*}
\end{linenomath}
The starred indices ${a*,b*,c*}$ correspond to the values of $a,b,c$ that yield the maximal absolute value of the difference between $X$ and $Y$ at each value of the wavevector argument $\mathbf{p}=(p_x,p_z)$. The normalization limits $V_{2}(\mathbf{k},\mathbf{p},\mathbf{q})$ to the range $[-2,2]$. If the assumption justifying the replacement of the average of the product by the product of averages is valid, the measure $V_2$ should be much smaller than one. 

Fig.~\ref{fig:V2_28} shows $V_2$ for the weakly forced simulation (for the other two simulations, see Supplementary Figs.~\ref{fig:V2_27} and \ref{fig:V2_29}). In areas of the wavevector space where the interaction term is non-negligible (that is, for $\mathbf{p}\approx0$ and for $\mathbf{q}=\mathbf{p}-\mathbf{k}\approx0$, Fig.~\ref{fig:interaction-terms_28_climn=0.1}), the values of $V_2$ are large and close to their maximal absolute value of 2, and are thus not consistent with this second derivation. While derivations in the literature simply assume $T\to\infty$, it is far from obvious what integration time $T$ should be used in this test of the second derivation. Because our DNS are in a statistical steady state and we do not expect a significant trend in wave amplitudes, we chose a long integration time of 200 days. One might need to use a time that is a function of the wave periods appearing in the definition of $X$ and $Y$. Yet, our results indicate that the assumption of a slow amplitude does not seem to hold for our idealized 2D simulations in the context of this second derivation as well.

\begin{figure*}
\includegraphics[width=\textwidth]{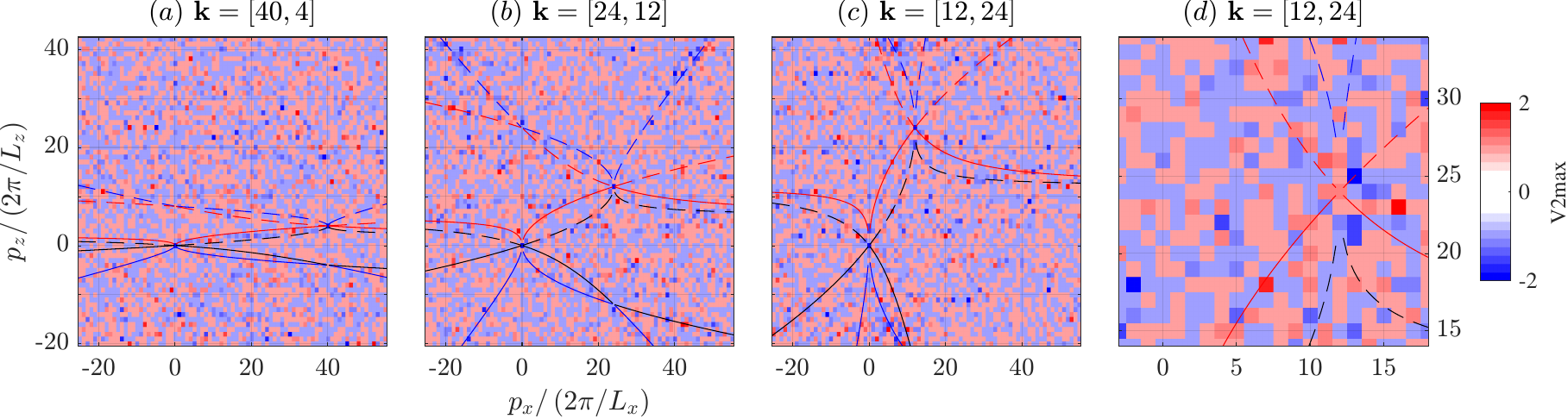}
\caption{Testing whether the three-product of wave amplitudes varies slowly, as quantified by criterion $V_2$, eq.~\eqref{eq:V2}, for the weak forcing case, for the wavevectors denoted in the titles of each panel.}
\label{fig:V2_28}
\end{figure*}

\subsection{Examining a third derivation of the resonant interactions approximation}
\label{sec:examining-third-derivation}

The commonly used derivation of the RWIA (Appendix~\ref{sec:app-ensemble-gaussian}) considers the equation for the time derivative of the triple product appearing in the energy equation \eqref{eq:energy-equation}, where the RHS of this equation contains a four-product of the wave amplitudes. See, for example, \cite{Lvov-Polzin-Yokoyama-2012:resonant} for an application of the weak wave turbulence's kinetic equation to internal waves based on a canonical Hamiltonian approach, or \cite{Nazarenko-2011:wave} where this derivation is applied to the generating function. One then applies an ensemble average, uses a Gaussian decomposition of the four-product, assumes a spatial translation invariance, and assumes that right- and left-propagating waves are uncorrelated. The equation describing the dynamics of the ensemble-averaged energy spectrum $E_{a,\mathbf{k}}(t)=\langle{}A_{a,\mathbf{k}}^*A_{a,\mathbf{k}}\rangle/V$, where $\langle\cdot\rangle$ denotes an ensemble average and $V$ is the volume, takes the following form, neglecting forcing and dissipation,
\begin{align}
\frac{\partial}{\partial t}E_{a}(\mathbf{k},t)
 &= \frac{1}{4(2\pi)^4}Re\Bigg(
\int\limits_{-\infty}^{\infty}
d^{2}\mathbf{q}d^{2}\mathbf{p}\delta\left(  \mathbf{k-p-q}\right)
{\displaystyle\sum\limits_{b,c}}
\mathsf{M}_{abc}\left(  \mathbf{k,p,q}\right)
\nonumber \\
&\times\int\limits_0^t dt'
e^{i\left(a\omega_{\mathbf{k}}-b\omega_{\mathbf{p}}-c\omega_{\mathbf{q}}\right)(t-t')}\mathcal{G}(\mathbf{k},\mathbf{p},\mathbf{q},a,b,c,t') dt'
\Bigg),
 \label{eq:enegry-equation-using-G}
\end{align}
where,
\begin{linenomath}
\begin{align}
    &\mathcal{G}(\mathbf{k},\mathbf{p},\mathbf{q},a,b,c,t')=
    \Big[\mathsf{M}_{abc}\left(\mathbf{k},\mathbf{p},\mathbf{q}\right)
 E_{b}(\mathbf{p},t') E_{c}(\mathbf{q},t')\nonumber
\nonumber\\
 &-\mathsf{M}_{bac}\left(\mathbf{p,k,q}\right)
 E_{a}(\mathbf{k},t')E_{c}(\mathbf{q},t')
-\mathsf{M}_{cab}\left(\mathbf{q,k,p}\right)
E_{a}(\mathbf{k},t')E_{b}(\mathbf{p},t')\Big].
\label{eq:define-G-paper}
\end{align}
\end{linenomath}
For more details and the derivation of $\mathcal{G}$, see Appendix \ref{sec:app-ensemble-gaussian}. The RHS of the energy equation then contains a time integral over products of the energy multiplied by $e^{i\Delta\omega t}$ (Eq.~\ref{eq:enegry-equation-using-G}). This integral is simplified \cite[e.g.,][]{Caillol-Zeitlin-2000:kinetic, Lvov-Polzin-Yokoyama-2012:resonant}, assuming that the ensemble-averaged energy spectrum varies over a timescale that is longer than the wave periods. This is assumed to lead to an integral over $e^{i\Delta\omega t}$ and thus to a delta-function factor $\delta({\Delta\omega})$, representing the resonance condition, leading to the final kinetic equation. 

However, in order to consider the time integral of Eq.~\ref{eq:enegry-equation-using-G} as the product of two integrals, a more proper assumption is that the timescale of the energy spectrum variation (and thereby of $\mathcal{G}$) should be much larger than the linear timescale of the oscillating term, $2\pi/\Delta\omega$. Letting $T_E$ be the timescale over which the ensemble-averaged energy varies, the condition for this third derivation to be valid is,
\begin{linenomath}
\begin{align}
   V_3(\mathbf{k},\mathbf{p},a,b,c)=\frac{2\pi}{\Delta\omega(\mathbf{k},\mathbf{p},\mathbf{k-p},a,b,c)}\times\frac{1}{T_E} \ll 1.
   \label{eq:V3}
\end{align}
\end{linenomath}
Given that, by definition, $\Delta\omega=0$ on resonance curves and that this linear time therefore diverges there, the time scale for energy variations, $T_E$, cannot be longer than the linear time scale on the resonance curves and therefore the condition in eq.~\eqref{eq:V3} cannot be satisfied for strictly resonant interactions. Supplementary Fig.~\ref{fig:V3} shows $V_3$, using a somewhat arbitrary $T_E=1$ day timescale, demonstrating that $V_3$ is also not small \textit{near} the resonance curves, as expected. If significant interactions occur \textit{near} resonance curves, rather than only \textit{on} the curves, as has been suggested \cite[][]{Lvov-Lvov-Newell-et-al-1997:statistical, Lvov-Polzin-Yokoyama-2012:resonant}, this condition needs to be satisfied at least near the curves, even if it cannot be satisfied on the curves, for this derivation to be self-consistent. 

\section{Discussion and conclusions}
\label{sec:discussion}

Numerous previous studies applied the resonant wave interaction approximation (RWIA) and the weak wave turbulence kinetic equation to internal waves \cite[e.g.,][]{Mccomas-Bretherton-1977:resonant, Caillol-Zeitlin-2000:kinetic, Kenyon-1968:wave}. The weak interaction hypothesis behind the kinetic equation was claimed to be inconsistent with observations \cite[][]{Holloway-1980:oceanic}, although the claim was disputed by more recent work \citep{Dematteis-Le-Pollmann-et-al-2024:interacting}, while the self-consistency of the kinetic equation was carefully examined by \cite[e.g.,][]{Lvov-Polzin-Yokoyama-2012:resonant}. Yet, to the best of our knowledge, no prior effort has examined whether the nonlinear dynamics due to the interaction of internal waves are consistent with the assumptions underlying the derivation of the RWIA. Equally important, previous studies have not examined the entire range of resolved wavenumbers to test if all significant wave-wave interactions satisfy the resonance constraint. We note that \cite{Pan-Arbic-Nelson-et-al-2020:numerical} did examine a few specific cross-sections in wavenumber space to find that some resonant interactions were significant, but did not consider all such resolved interactions as we do here for a highly idealized non-rotating 2D case. 

Given the known differences between 3D and 2D systems and the obvious importance of the Coriolis force in the ocean, it cannot be claimed that our results apply to the ocean's internal waves. However, we are not the first to consider the weak wave turbulence formalism applied to 2D internal waves, as was done, for example, by \citet{Ripa-1981:theory, Daubner-Zeitlin-1996:stationary, Caillol-Zeitlin-2000:kinetic}, and more recently by \citet{Shavit-Buehler-Shatah-2025:turbulent, Labarre-Shavit-2026:distinguished}, who compared a 2D simulated spectrum to the predictions of the kinetic equation. In this context, we set out to examine the consistency of the RWIA within the limited framework of idealized, 2D, double periodic, non-rotating simulations. The weak forcing regime we study is consistent with that where the RWIA should hold. Yet another important validity requirement for this approximation, of an infinite domain \cite[][]{Nazarenko-2011:wave}, cannot be satisfied in numerical simulations such as those used here.

We find that the dominant interactions in our direct numerical simulations are not necessarily resonant, and that resonant interactions are mostly not significant in terms of their contributions to energy transfers between different wavevectors. In order to attempt to rationalize this, we examine three derivations of the RWIA. None of the three derivations considered is consistent with our direct numerical simulations because the assumption that the wave amplitudes or energy vary slower than the relevant linear wave time scales is violated even for our weakest forced simulation, and the timescale over which the slow amplitude changes at a given wavevector is therefore \textit{not} much larger than the wave period, as required by the derivations. The rapidly varying amplitudes in our simulations, despite the very weak forcing and nonlinear interactions, suggest that it may not be completely justified to think of the fluid motion as interacting linear internal waves, as such a description implies slowly varying wave amplitudes. We conclude that the RWIA does not seem to be consistent with the interactions seen in our idealized direct numerical simulations.

Our objective in this work was to test the RWIA in the particular application to 2D non-rotating, double-periodic internal waves. Whether these results generalize is not obvious and beyond the scope of this work. Our simulations are meant to examine the RWIA in the simplest configuration. Many of our results, while carefully derived, rely on the adequacy and relevance of the direct numerical simulations employed here. The idealizations employed here, including the double periodic 2D domain, are similar to those used in recent simulations of internal ocean waves \cite[][]{Sugiyama-Niwa-Hibiya-2009:numerically, Eden-Chouksey-Olbers-2019:mixed, Chen-Chen-Liu-et-al-2019:can}. Yet, the simulations span a range of no more than 2 orders of magnitude of wavevectors that are not dominated by the numerical model viscosity. We do not include the Coriolis force nor the abundant near-inertial motions generated in the ocean by the winds at the surface and by the interaction of internal tides with topography. Our use of 2D vs.\ 3D simulations means that the number of resonant interactions in any range of wave frequencies is lower, potentially weakening the resonant interactions \cite[][]{Lvov-Nazarenko-Pokorni-2006:discreteness}, and we further discuss this in the following paragraph. 

The use of 2D simulations also implies that we do not resolve vertical vorticity due to the horizontal shear of the horizontal velocity (vortical modes such as quasigeostrophic eddies), which was found to play an important role in scattering internal wave energy \cite[][]{Cox-Kafiabad-Vanneste-2023:inertia, Savva-Kafiabad-Vanneste-2021:inertia, Kafiabad-Savva-Vanneste-2019:diffusion, Young-2021:inertia, Barkan-Srinivasan-McWilliams-2024:eddy}. The system of equations we explore \eqref{eq:momentum-equations-dim} allows a steady state solution of the form $u=u(z)$, $w=0$, $b=0$. In a weakly nonlinear system, such a solution may play the role of a slowly varying vortical mode that exists in a 3D system. This mode is also possible to interpret as the ``condensates'' that were studied by \cite{Korotkevich-Nazarenko-Pan-et-al-2024:non}, who suggested a modification of the spectral diffusion equation that includes the effects of such condensates. Our simulations indeed do develop an $x$-independent $u(z)$ mode. The equations we use \eqref{eq:A-equation} include this mode and its interaction with other waves. Yet, to verify that this mode is not the reason for the non-resonant interactions we find to be dominant, we ran another simulation where $\mathbf{u}(k_x=0)$ and $\mathbf{u}(k_z=0)$ are both set to zero. The results in Fig.~\ref{fig:interaction-terms_3_climn=0.1} show that this does not affect the main result that non-resonant interactions are dominant. We also note that in the case of \cite{Korotkevich-Nazarenko-Pan-et-al-2024:non} the condensates seem to contain a significant part of the energy, while this is not the case in our simulations (see the spectra in Fig.~\ref{fig:characterization}). Our use of a constant buoyancy frequency eliminates some nonlinear interactions that can occur only with non-uniform stratification \cite[][]{Sutherland-2016:excitation}. The nonlinearity parameter in our simulations ($\epsilon$) seems exceedingly small, yet it would have been reassuring if we could identify an even smaller value where only resonant interactions dominate. Unfortunately, further reducing the nonlinearity by reducing the forcing amplitude means that the small numerical dissipation used for numerical reasons prevents the wave energy from spreading to large wavevectors, making the inertial range too small to be meaningful.

Another concern that may be raised regarding our analysis is that the simulations do not allow for a sufficient number of resonant interactions to be excited. The RWIA involves taking two limits: the system size going to infinity and the nonlinearity being very weak \cite[e.g.,][]{Nazarenko-2011:wave}. In numerical simulations, the resolved wavenumbers are discrete and cannot be expected to lie exactly on the resonant curves. This might suggest that a discrete simulation cannot be used to test the RWIA. Of course, linear dissipation, which is included in our numerical model as diffusion and viscosity, will lead to broadening. In addition, it was shown that nonlinear interactions lead to the broadening of the resonance curves to include near-resonant interactions \cite[][]{Lvov-Lvov-Newell-et-al-1997:statistical, Lvov-Polzin-Yokoyama-2012:resonant, Pan-Yue-2017:understanding, Zhang-Pan-2022:numerical} within the RWIA. This broadening implies that under the assumptions of the RWIA, near-resonance interactions are expected to dominate the energy transfer and allow an energy cascade to develop \cite[][]{Lvov-Nazarenko-Pokorni-2006:discreteness}. It is difficult to estimate a priori the specific broadening for a given strength of the nonlinear interactions and numerical resolution. It is also not known theoretically what number of near-resonant interactions is required for weak wave turbulence to be well-represented on a finite grid. In order to address these issues, we invoke the estimates of the nonlinear broadening of resonance curves estimated by \cite{Lvov-Polzin-Yokoyama-2012:resonant}. They find that a Lorentzian broadening, with the width set to 0.5 times the sum of frequencies of the three interacting waves, leads to a self-consistent Boltzmann ratio. We analyzed their suggested Lorentzian weight applied to near-resonance wavenumbers and considered any triad whose weight is larger than 0.75 to be an activated near-resonance interaction. For the three wavenumbers considered in our different figures, this broadening leads to 35\%--75\% of the wavevectors being activated(!). It seems, therefore, that our simulations contain a sufficient number of activated interactions within the near-resonance range of the RWIA for a realistic level of nonlinearity in observed ocean internal waves. Another mechanism that leads to the broadening of resonances and to the activation of near-resonant interactions was discussed by \cite{Annenkov-Shrira-2006:role} and \cite{Janssen-2003:nonlinear}. We note that there are alternative estimates of the nonlinear broadening \cite[][]{Pan-Yue-2017:understanding, Zhang-Pan-2022:numerical}. They point out that the limit of infinite integration time that leads to the delta function corresponding to the resonant frequencies (see definition and limit of the function $\delta_T\left(\Delta\omega \right)$ in eqs.~\ref{eq:delta_T} and \ref{eq:sinc-limit}) is not physical, as the wave amplitudes would be expected to vary over such a long time due to nonlinear interactions. As a result, they consider a finite time limit in which this function maintains a finite width and leads to the inclusion of near-resonance interactions to be more realistic. They conclude that this broadening justifies the application of the kinetic equation to the analysis of a finite-resolution DNS as we do here. A related broadened-resonance analysis and a resulting generalized kinetic equation were applied to the parametric subharmonic instability by \cite{Onuki-Hibiya-2019:parametric}. Any numerical simulations cannot strictly satisfy the large-box limit required for the kinetic equation to hold in the sense that this limit allows for enough strict resonances to exist \cite[][]{Nazarenko-2011:wave}. Yet, the above discussion of processes leading to the broadening of resonances means that even at a finite system size, we expect to have enough activated resonances. While this discussion suggests that the discreteness and the finite domain size of the numerical simulation may not be major issues for our results, it would be good to use the same analysis approach with numerical simulations that use a much higher resolution and a larger domain size.

A final caveat is that our analysis of the assumptions used to derive the RWIA is based on time averaging, while the standard derivation (Appendix~\ref{sec:app-ensemble-gaussian}) relies on ensemble averaging and the Gaussian decomposition of four-products of wave amplitudes. We have also not addressed a recent alternative derivation of the kinetic equation based on large deviations theory \cite[][]{Guioth-Bouchet-Eyink-2022:path}.

It is important to note that while we find issues with the assumptions used to derive the kinetic equation when tested in our highly idealized framework, the kinetic equation has been used successfully for explaining observed internal ocean characteristics. \cite{Onuki-Hibiya-2018:decay} and \cite{Olbers-Pollmann-Eden-2020:psi}, for example, showed that the kinetic equation can correctly estimate the decay rates of internal waves via the parametric subharmonic instability and the transfer of tidal energy to the wave continuum. Similarly, \cite{Lvov-Tabak-2001:hamiltonian} showed that the kinetic equation, based on Hamiltonian formalism, results in a GM-like energy spectrum. \cite{Pan-Arbic-Nelson-et-al-2020:numerical} showed that in a high-resolution realistic simulation, scale-separated interactions such as induced diffusion are the most important mechanisms for the formation of the power-law spectrum. This last result is consistent with our finding of strong interactions with low wave numbers in Fig.~\ref{fig:interaction-terms_28_climn=0.1}, although we note that non-resonant interactions with low wave numbers are also important in our simulations. The kinetic equation was also used to study energy fluxes in stationary and non-stationary spectra \cite[][]{Dematteis-Lvov-2023:structure, Dematteis-Lvov-2021:downscale}.

We attempted to make the case above that none of the idealizations used here, nor our model limitations, would necessarily mean that the RWIA should not have applied to our simulation results, given that the measure of the nonlinearity ($\epsilon$ above) is very small for all our simulations, and that this weak nonlinearity is the basis for the slowness assumption used in all derivations of the RWIA. Our findings, while perhaps of limited applicability due to the limited numerical resolution and highly idealized setting, call for further testing of the assumptions underlying the RWIA in more sophisticated and realistic simulations following the methodology outlined here.

\textbf{Acknowledgments.} We thank Kraig Winters for making the code of flow\_solve available, and Yulin Pan, Nicolas Grisouard, and Oliver Buhler for helpful comments and discussions. ET thanks the Weizmann Institute for its hospitality during parts of this work and has been funded by the NSF Climate Dynamics Program (joint NSF/NERC) grant AGS-1924538. GB thanks Ester Levi for a lifelong inspiration.






\newpage
\bibliography{all,mypapers}

\newpage
\appendix
\renewcommand{\thetable}{A-\arabic{table}}

\setcounter{secnumdepth}{2}
\setcounter{equation}{0}
\renewcommand{\theequation}{\thesection-\arabic{equation}}
\setcounter{table}{0}
\renewcommand{\thetable}{A-\arabic{table}}

\section{Nondimensionalization}
\label{sec:appendix-non-dimensionalization}

\allowdisplaybreaks
The dimensional momentum equations in terms of dimensional variables denoted with $\tilde{(\cdot)}$ are
\begin{linenomath}
\begin{align}
\label{eq:momentum-equations-dim}
   \widetilde{u}_{\widetilde{t}}+\mathbf{\widetilde{u}}\cdot
   \widetilde{\nabla}\widetilde{u} 
   & =-\frac{1}{\rho_{0}}\widetilde
{p}_{\widetilde{x}}+\widetilde{\nu}_{x}\frac{\partial^{2}}{\partial
\widetilde{x}^{2}}\widetilde{u}+\widetilde{\nu}_{z}\frac{\partial^{2}%
}{\partial\widetilde{z}^{2}}\widetilde{u}, 
\nonumber\\
\widetilde{w}_{\widetilde{t}}+\mathbf{\widetilde{u}}\cdot
\widetilde{\nabla}\widetilde{w} &  =-\frac{1}{\rho_{0}}\widetilde
{p}_{\widetilde{z}}+\widetilde{b}+\widetilde{\nu}_{x}\frac{\partial^{2}%
}{\partial\widetilde{x}^{2}}\widetilde{w}+\widetilde{\nu}_{z}\frac
{\partial^{2}}{\partial\widetilde{z}^{2}}\widetilde{w}.
\end{align}
\end{linenomath}
Here, $\widetilde{\mathbf{x}}=[\widetilde{x},\widetilde{z}]$ is the [horizontal, vertical] dimensional position vector, and the corresponding velocity vector is $\mathbf{\widetilde{u}}=[\widetilde{u},\widetilde{w}]$. The equation for the buoyancy $\widetilde{b}=-g\widetilde{\rho}/\rho_{0}$, where $\widetilde{\rho}$ is the dimensional density, $g=9.8$ m\,s$^{-2}$ is gravitational acceleration, and $\rho_0=1000$ kg/m$^3$, is
\begin{linenomath}
\begin{align}
\label{eq:buoyancy-equation-dim}
\widetilde{b}_{\widetilde{t}}+\mathbf{\widetilde{u}}\cdot
\widetilde{\nabla}\widetilde{b}+\widetilde{w}\overline{\widetilde{b}}_{\widetilde{z}}=
\widetilde{\kappa}_{x}\frac{\partial^{2}}{\partial\widetilde{x}^{2}%
}\widetilde{b}+\widetilde{\kappa}_{z}\frac{\partial^{2}}{\partial\widetilde
{z}^{2}}\widetilde{b}+\widetilde{f}^{(b)}\left(  \mathbf{\widetilde
{x}},\widetilde{t}\right).
\end{align}
\end{linenomath}
The incompressibility condition implies,
\begin{linenomath}
\begin{align}
\widetilde{u}_{\widetilde{x}}+\widetilde{w}_{\widetilde{z}}=0.
\end{align}
\end{linenomath}
In our direct numerical simulations, we impose a constant mean stratification, which dictates the Brunt–Väisälä buoyancy frequency, $N$, as $\overline{\widetilde{b}}_{\widetilde{z}}={N}^{2}$.

In order to nondimensionalize the equations, we choose the following length, time, and velocity scales,
\begin{linenomath}
\begin{equation}
\widetilde{x}=xL_x,\ 
\widetilde{z}=zL_z,\ 
\widetilde{t}=t/N,\ 
\widetilde{u}=uu_{0},\ 
\widetilde{w}=ww_{0}=w\,u_{0}\frac{L_z}{L_x},
\end{equation}
\end{linenomath}
and the nondimensionalization of the pressure and buoyancy are given by
\begin{linenomath}
\begin{align}
\widetilde{p} &  = pp_{0},\ \text{where }p_{0}=\rho_{0}L_x N u_{0},\nonumber \\
\widetilde{b} &  =bb_{0}, \text{where }b_{0}=\frac{L_z}{L_x}u_{0}N.
\end{align}
\end{linenomath}
Applying this to the momentum equations, one finds
\begin{linenomath}
\begin{align}
   \label{eq:nondim-SI}
   u_{t}+\epsilon\mathbf{u}\cdot\nabla u &  =-p_{x}+\left(
   \nu_{x}\frac{\partial^{2}u}{\partial x^{2}}
   +\nu_{z}\frac{\partial^{2} u}{\partial z^{2}}\right),
   \nonumber  \\
   w_{t}+\epsilon\mathbf{u}\cdot\nabla w &  =-p_{z}+b
   +\left(\nu_{x}\frac{\partial^{2}w}{\partial x^{2}}+\nu_{z}\frac{\partial^{2}
   w}{\partial z^{2}}\right),
   \nonumber \\
   b_{t}+\epsilon\mathbf{u}\cdot\nabla b+w &  
   =\left(  \kappa_{x}\frac{\partial^{2}b}{\partial x^{2}}+\kappa_{z}
   \frac{\partial^{2} b}{\partial z^{2}}\right)  
   +f^{(b)}\left(\mathbf{x},t\right),
   \nonumber \\
   u_{x}+w_{z}&=0,
\end{align}
\end{linenomath}
where we defined the components of the dimensionless anisotropic eddy viscosity $\nu_{x}=\widetilde{\nu}_{x}/{NL_x^{2}}$ and $\nu_{z}=\widetilde{\nu}_{z}/{NL_z^{2}}$, and the components of the dimensionless anisotropic eddy diffusivity, $\kappa
_{x}=\widetilde{\kappa}_{x}/{NL_x^{2}};\kappa_{z}=\widetilde{\kappa}
_{z}/{NL_z^{2}}$. The dimensionless forcing term is,
\begin{linenomath}
\begin{align*}
f^{(b)}\left(  \mathbf{x},t\right)  =\frac{1}{Nb_{0}}\widetilde
{f}^{(b)}\left(  \mathbf{\widetilde{x}},\widetilde{t}\right).
\end{align*}
\end{linenomath}
The nondimensional parameter multiplying the nonlinear terms is,
\begin{linenomath}
\begin{align}
\epsilon={u_{0}}/{NL_x}. 
   \label{eq:epsilon-definition}
\end{align}
\end{linenomath}
The velocity scale appearing here is calculated as the RMS of horizontal velocity for each run, while $L_x$ and $L_z$ are the horizontal and vertical basins' extents, respectively.

\section{Derivation of the energy equation in spectral space}
\label{sec:app-derivation-energy-equation}

Based on the nondimensional equations derived in the previous section (eqs.~\ref{eq:nondim-SI}) we introduce a stream function $\psi$ such that $u=\psi_{z}$, and $w=-\psi_{x}$ so that the vorticity is $\zeta=u_{z}-w_{x}=\nabla^{2}\psi$. One can take the curl of the momentum equations to derive a vorticity equation and rewrite the buoyancy equation to find
\begin{linenomath}
\begin{align}
\label{eq:vorticity-buoyancy-nondim}
\zeta_{t}-(\nu_x\zeta_{xx}+\nu_z\zeta_{zz})+b_{x} 
& =\epsilon J(\psi,\zeta),
\nonumber\\
b_{t}-(\kappa_xb_{xx}+\kappa_zb_{zz})-\psi_{x} 
& =\epsilon J(\psi,b)+f^{b}(\mathbf{x},t).
\end{align}
\end{linenomath}
Here, $J(a,b)\equiv{}a_xb_z-a_zb_x$. Fourier transforming eqs.~\eqref{eq:vorticity-buoyancy-nondim} yields
\begin{linenomath}
\begin{align}
   \label{eq:vorticity-buoyancy-nondim-FT}
   &\dot{\zeta}_{\mathbf{k}}
   +(\nu_{x}k_{x}^{2}+\nu_{z}k_{z}^{2})\zeta_{\mathbf{k}}
  +ik_{x}b_{\mathbf{k}} 
  =\epsilon(2\pi)^{-2}\int_{-\infty}^{\infty}\!\!\!d^{2}\mathbf{p}\,d^{2}\mathbf{q}\,
  \delta(\mathbf{p}+\mathbf{q}-\mathbf{k})(p_{x}q_{z}-p_{z}q_{x})p^{-2}
  \zeta_{\mathbf{p}}\zeta_{\mathbf{q}}
   \nonumber\\
   &\dot{b}_{\mathbf{k}}+(\kappa_x k_x^{2}+\kappa_zk_z^2)b_{\mathbf{k}}+ik_{x}\zeta_{\mathbf{k}}k^{-2} 
   \nonumber \\
   &=\epsilon(2\pi)^{-2}\int_{-\infty}^{\infty}\!\!\!d^{2}\mathbf{p}\,d^{2}\mathbf{q}\,
   \delta(\mathbf{p}+\mathbf{q}-\mathbf{k})(p_{x}q_{z}-p_{z}q_{z})p^{-2}
   \zeta_{\mathbf{p}}b_{\mathbf{q}} + f^{b}(\mathbf{k},t).
\end{align}
\end{linenomath}
Here, $f^{b}(\mathbf{k},t)$ is the Fourier transform of the forcing term of the buoyancy equation. Next, we define a new variable \citep[e.g.,][]{Carnevale-1981:statistical}, $A_{a,\mathbf{k}}$, which is a wave amplitude,
\begin{linenomath}
\begin{align*}
A_{a,\mathbf{k}}=\frac{1}{2}(\zeta_{\mathbf{k}}/k+ab_{\mathbf{k}}),
\end{align*}
\end{linenomath}
where $a=\pm1$ represents right- and left-propagating waves. The equation of motion for the wave amplitude in the spectral domain takes the form,
\begin{linenomath}
\begin{align}
\dot{A}_{a,\mathbf{k}}+\mathsf{L}_{ab}(\mathbf{k})A_{b,\mathbf{k}} 
 =\epsilon{\sum_{\mathbf{p},\mathbf{q}}}{\sum\limits_{b,c}}
 \delta\left(\mathbf{k}-\mathbf{p}-\mathbf{q}\right)
 \mathsf{M}_{abc}(\mathbf{k},\mathbf{p},\mathbf{q})A_{b,\mathbf{p}}A_{c,\mathbf{q}}+F_{a}(\mathbf{k}),
 \label{eq:dot_A^s_k}
\end{align}
\end{linenomath}
where the interaction coefficient $\mathsf{M}_{\pm}$ and the linear operator $\mathsf{L}$, that represents both the wave and the dissipation, are
\begin{linenomath}
\begin{align}
\mathsf{M}_{\pm{}}(\mathbf{k},\mathbf{p},\mathbf{q}) 
   &=\frac{p_{x}q_{z}-p_{z}q_{x}}{kpq}
   \left(\begin{array}{cc}
   q^{2}-p^{2}\pm k(q-p) & q^{2}-p^{2}\mp k(q+p)\\
   q^{2}-p^{2}\pm k(q+p) & q^{2}-p^{2}\mp k(q-p)
   \end{array}\right),  
   \nonumber \\
   \mathsf{L_d}(\mathbf{k})&=\left(
   \begin{array}[c]{cc}
   \frac{1}{2}\left(  \nu_{x}+\kappa_{x}\right)  k_{x}^{2}+\frac{1}{2}\left(
   \nu_{z}+\kappa_{z}\right)  k_{z}^{2} & \frac{1}{2}\left(
   \nu_{x}-\kappa_{x}\right)  k_{x}^{2}+\frac{1}{2}\left(  \nu_{z}-\kappa_{z}\right)  k_{z}^{2}\\
   \frac{1}{2}\left(  \nu_{x}-\kappa_{x}\right)  k_{x}^{2}+\frac{1}{2}\left(
   \nu_{z}-\kappa_{z}\right)  k_{z}^{2} & \frac{1}{2}\left(  \nu_{x}+\kappa_{x}\right)  k_{x}^{2}+\frac{1}{2}\left(  \nu_{z}+\kappa_{z}\right)  k_{z}^{2}
   \end{array} \right), 
   \nonumber \\
   \mathsf{L_w}(\mathbf{k})&=\left(
   \begin{array}[c]{cc}
     +i{k_{x}}/{k} & 0 \\
     0 & -i{k_{x}}/{k}
   \end{array}\right),\quad  
   \mathsf{L}=\mathsf{L_w}+\mathsf{L_d}.
   \label{eq:M-and-L-operators}
 \end{align}
\end{linenomath}
The linear wave solution, ignoring all forcing, friction, and nonlinear terms is ${A}_{a,\mathbf{k}}={A}_{a,\mathbf{k}}(0)e^{-ia(k_x/k)t}$ and the corresponding nondimensional frequency is given by $-ia(k_x/k)$. The forcing term $F_{s}(\mathbf{k},t)=sf^{b}(\mathbf{k},t)/2$ in the amplitude equation~\eqref{eq:dot_A^s_k} is due to the forcing terms appearing in equations~\eqref{eq:vorticity-buoyancy-nondim-FT}. The energy equation in terms of the wave amplitudes is,
\begin{linenomath}
\begin{align}
  \label{eq:energy_equation_appendix}
  &\frac{\partial}{\partial t}
   \left(A_{a,\mathbf{k}}A_{a,\mathbf{k}}^{\ast}\right)   
  =-2Re\left({\sum\limits_{b}}
   L_{ab}A_{b,\mathbf{k}}A_{a,\mathbf{k}}^{\ast}\right) 
   +2Re\left(F_{a}^{\ast}\left(  \mathbf{k}\right)  
   A_{a,\mathbf{k}}\right)
\nonumber \\
   &+\epsilon2Re\left(\frac{1}{4\left(  2\pi\right)^{2}}
   {\int\limits_{-\infty}^{\infty}}
   d^{2}\mathbf{q}d^{2}\mathbf{p}
   \delta\left(\mathbf{k}-\mathbf{p}-\mathbf{q}\right) 
   {\sum\limits_{b,c}}
   \mathsf{M}_{abc}\left(\mathbf{k,p,q}\right)
   A_{b,\mathbf{p}}A_{c,\mathbf{q}}A_{a,\mathbf{k}}^{\ast}\right).
\end{align}
\end{linenomath}

\section{First derivation: a two-timescale perturbative approach}
\label{sec:app-2-time-scale}

In the first derivation considered here of the resonant condition on the frequencies of interacting waves, one uses a multiple timescale perturbative approach. See \cite{Benney-Saffman-1966:nonlinear}, or as outlined in the context of internal waves in, e.g., \cite{Caillol-Zeitlin-2000:kinetic}, with the general multiple timescale perturbative method described in Section 11.2 of \cite{Bender-Orszag-1978:advanced}. We start by assuming that the solution can be expanded in the small parameter, $\epsilon$, namely,
\begin{linenomath}
\begin{align}
A_{s,\mathbf{k}}=A_{s,\mathbf{k,}0}+\epsilon A_{s,\mathbf{k,}1}+\epsilon
^{2}A_{s,\mathbf{k,}2}+\cdots = %
{\displaystyle\sum\limits_{m=0}^{\infty}}
\epsilon^{m}A_{s,\mathbf{k,}m}.
\end{align}
\end{linenomath}
We also assume a timescale separation, namely, we define a slow time $\tau=\epsilon t$ and the full time derivative is assumed to be of the form $\frac{d}{dt}%
=\frac{\partial}{\partial t}+\epsilon\frac{\partial}{\partial\tau}$. By substituting the expansion above in the amplitude equation (eq.~\eqref{eq:A-equation} in the main text), we find,
\begin{linenomath}
\begin{align}
&\frac{\partial}{\partial t}
{\displaystyle\sum\limits_{m=0}^{\infty}}
\epsilon^{m}A_{a,\mathbf{k,}m}+\frac{\partial}{\partial\tau}
{\displaystyle\sum\limits_{m=0}^{\infty}}
\epsilon^{m+1}A_{a,\mathbf{k,}m}+L_{ar}
{\displaystyle\sum\limits_{m=0}^{\infty}}
\epsilon^{m}A_{r,\mathbf{k,}m}+F_{a}\left(  \mathbf{k}\right)\nonumber\\
&=\frac{\epsilon}{4\left(2\pi\right)^{2}}
{\displaystyle\int\limits_{-\infty}^{\infty}}
d^{2}\mathbf{q}d^{2}\mathbf{p}
\delta\left(\mathbf{k-p-q}\right)
\mathsf{M}_{abc}\left(\mathbf{k,p,q}\right)
{\displaystyle\sum\limits_{m=0}^{\infty}}
\epsilon^{m}A_{b,\mathbf{p},m}
{\displaystyle\sum\limits_{m'=0}^{\infty}}
\epsilon^{m'}A_{c,\mathbf{q},m'}.
\end{align}
\end{linenomath}
The zeroth order variable simply satisfies the linear equation,
\begin{linenomath}
\begin{align}
   \frac{\partial}{\partial t}
   A_{a,\mathbf{k,}0}+L_{ab}A_{b,\mathbf{k,}0}
   =F_{a}\left(\mathbf{k}\right),
\end{align}
\end{linenomath}
with a solution for $\mathbf{k}$ away from the forced wavevectors of the form (neglecting dissipation and diffusion),
\begin{linenomath}
\begin{align}
  A_{a,\mathbf{k,}0} =\hat{A}_{a,\mathbf{k}}\left(  \tau\right)
  e^{-ia\omega_\mathbf{k}t}.
\end{align}
\end{linenomath}
The equation for the first order is
\begin{linenomath}
\begin{align}
&\frac{\partial}{\partial t}
A_{a,\mathbf{k,}1}+L_{ab}A_{b,\mathbf{k,}1}  
=-\frac{\partial}{\partial\tau}
A_{a,\mathbf{k,}0} \nonumber \\
 & +\frac{1}{4\left(2\pi\right)^{2}}
{\displaystyle\int\limits_{-\infty}^{\infty}}
d^{2}\mathbf{q}d^{2}\mathbf{p}
\delta\left(  \mathbf{k-p-q}\right)
\mathsf{M}_{abc}\left(\mathbf{k,p,q}\right)  A_{b,\mathbf{p,}0}A_{c,\mathbf{q,}0}.
\end{align}
\end{linenomath}
Substituting the solution of the zeroth order, the RHS becomes
\begin{linenomath}
\begin{align}
&e^{-ia\omega_\mathbf{k}t}\Bigg( 
-\frac{\partial}{\partial\tau}\hat
{A}_{a,\mathbf{k}}\left(  \tau\right)
\nonumber \\
&+\frac{1}{4\left( 2\pi\right)^{2}}{\int\limits_{-\infty}^{\infty}}
d^{2}\mathbf{q}d^{2}\mathbf{p}\delta\left(\mathbf{k-p-q}\right)
\mathsf{M}_{abc}\left(\mathbf{k,p,q}\right)  
\hat{A}_{b,\mathbf{p}}\left(\tau\right)  
\hat{A}_{c,\mathbf{q}}\left(\tau\right) 
e^{i\Delta\omega(\mathbf{k},\mathbf{p},\mathbf{q},a,b,c)t}\Bigg).
\label{eq:two-timescale-consistency-condition}
\end{align}
\end{linenomath}
This is proportional to $e^{-ia\omega(\mathbf{k})t}$, which is the solution to the LHS operator and, thus, may result in a secular term that grows linearly in $t$ and invalidates the perturbation expansion for large $t$ by making the $O(\epsilon)$ term larger than the zeroth order term. To prevent a secular term in the solution for the first-order variables, we demand that the expression in the parentheses, which is the coefficient of the term that leads to a secular term, vanishes. Because the first term in the parentheses is only a function of $\tau$, this implies that the second term, the integral, should not be a function of $t$ as well. Indeed, assuming a large time $t$, the rapidly oscillating exponent of $\Delta\omega$ will lead to the vanishing of the integral whenever $\Delta\omega\ne0$. This leads to the resonance condition,
\begin{linenomath}
\begin{align}
    \Delta\omega(\mathbf{k},\mathbf{p},\mathbf{q},a,b,c)
    =a\omega_\mathbf{k}-b\omega_\mathbf{p}-c\omega_\mathbf{q}=0,
    \label{eq:two-timescale-sum-freq-vanishes}
\end{align}
\end{linenomath}
also eliminating the $t$ dependence mentioned in the last sentence.

\section{Second derivation: assuming a slowly varying triple correlation}
\label{sec:app-slow-tripple-correlation}

Neglecting the viscosity and diffusion, the time-averaged energy equation \eqref{eq:energy_equation_appendix} in terms of the slow wave amplitudes becomes,
\begin{linenomath}
\begin{align}
   \label{eq:time_average_energy_equation}
   &\frac1T\int_0^T\frac{\partial}{\partial t}
   \left(A_{a,\mathbf{k}}A_{a,\mathbf{k}}^{\ast}\right)dt     
   =2Re\Bigg(\frac{\epsilon}{4\left(  2\pi\right)  ^{2}}
   {\displaystyle\int\limits_{-\infty}^{\infty}}
   d^{2}\mathbf{q}d^{2}\mathbf{p}
   \delta\left(  \mathbf{k-p-q}\right)
   {\displaystyle\sum\limits_{b,c}}
   \mathsf{M}_{abc}\left(\mathbf{k,p,q}\right)
   \nonumber \\
   & \times\frac{1}{T}\int_0^T
   \hat{A}_{a,\mathbf{k}}^{\ast}\hat{A}_{b,\mathbf{p}}\hat{A}_{c,\mathbf{q}}
    e^{i\Delta\omega(\mathbf{k},\mathbf{p},\mathbf{q},a,b,c)t}dt\Bigg)
  +2Re\left(\frac{1}{T}\int_0^TF_{a}^{\ast}\left(  \mathbf{k}\right)  
  e^{-ia\omega_{\mathbf{k}}t}\hat{A}_{a,\mathbf{k}}dt\right).
\end{align}
\end{linenomath}
The RHS involves the following integral,
\begin{linenomath}
\begin{align}
  \label{eq:Y}
  Y_{a,b,c}(\mathbf{k},\mathbf{p},\mathbf{q})\equiv Re \Bigg(\frac{1}{T} \int\limits_0^T
  \hat{A}_{a,\mathbf{k}}^{\ast}\hat{A}_{b,\mathbf{p}}\hat{A}_{c,\mathbf{q}} 
  e^{i\Delta\omega(\mathbf{k},\mathbf{p},\mathbf{q},a,b,c)t}dt\Bigg).  
\end{align}
\end{linenomath}
There are two possible paths to deriving the resonance condition leading to the kinetic equation starting from this equation. First, one assumes that the triple products of the slow amplitudes are uncorrelated with the deterministic phases; this reduces $Y_{a,b,c}(\mathbf{k},\mathbf{p},\mathbf{q})$ to,
\begin{linenomath}
\begin{align}
   \label{eq:X}
   X_{a,b,c}(\mathbf{k},\mathbf{p},\mathbf{q})&\equiv Re\Bigg(\frac{1}{T}\int\limits_0^T
   \hat{A}_{a,\mathbf{k}}^{\ast}\hat{A}_{b,\mathbf{p}} \hat{A}_{c,\mathbf{q}} dt\;\times\;
  \frac{1}{T}\int\limits_0^T
   e^{i\Delta\omega(\mathbf{k},\mathbf{p},\mathbf{q},a,b,c) t}dt\Bigg) 
   \nonumber\\
  &=Re\Bigg(\frac{1}{T}\int\limits_0^T
   \hat{A}_{a,\mathbf{k}}^{\ast}\hat{A}_{b,\mathbf{p}} \hat{A}_{c,\mathbf{q}}
   dt\;\times\;
 \frac{1}{T} 
 \delta_T\left(\Delta\omega(\mathbf{k},\mathbf{p},\mathbf{q},a,b,c)\right)\ \Bigg)
\end{align}
\end{linenomath}
where
\begin{align}
    \delta_T\equiv
   \frac{e^{i\Delta\omega(\mathbf{k},\mathbf{p},\mathbf{q},a,b,c)T}-1}
   {i\Delta\omega(\mathbf{k},\mathbf{p},\mathbf{q},a,b,c)}.
   \label{eq:delta_T}
\end{align}
In the limit of $T\to\infty$, the imaginary part of $\delta_T$ vanishes and we have 
\begin{linenomath}
    \begin{align}
    \lim_{T\to\infty} \delta_T\left(\Delta\omega(\mathbf{k},\mathbf{p},\mathbf{q},a,b,c)\right)
   &= \lim_{T\rightarrow\infty} \frac{\sin\left(
\Delta\omega(\mathbf{k},\mathbf{p},\mathbf{q},a,b,c)T\right)}
   {\Delta\omega(\mathbf{k},\mathbf{p},\mathbf{q},a,b,c)} 
   \nonumber \\
&=\pi \delta(\Delta\omega(\mathbf{k},\mathbf{p},\mathbf{q},a,b,c))
   \label{eq:sinc-limit}
\end{align}
\end{linenomath}
and the time-averaged product of three slow amplitudes is multiplied by a Dirac delta function of $\Delta\omega$, namely, we obtain the resonance condition on the frequencies of the interacting waves in \eqref{eq:time_average_energy_equation}.

The second path to derive the resonance condition starting from eq.~\eqref{eq:Y} involves the assumption that there is a time scale separation in the integral appearing in equation \eqref{eq:Y}. The first time scale arises from the oscillating exponent, 
\begin{equation}
  \label{eq:Tlinear}
  T_{\mathrm{linear}} =2\pi/\Delta\omega(\mathbf{k},\mathbf{p},\mathbf{q},a,b,c), 
\end{equation}
and the second time scale arises from the nonlinear interactions affecting the triple product there. For the time scale separation to work, the nonlinear time scale should be longer than the linear time scale. If there is such a time scale separation, the integral over the oscillating term may be taken separately, thereby leading to the resonance condition. We note, though, that such a separation cannot occur on the resonance curves, as the linear time scale diverges there, as further discussed in the main text.

\section{Derivation of the kinetic equation using an ensemble average and Gaussian approximation}
\label{sec:app-ensemble-gaussian}

\allowdisplaybreaks

For completeness, we provide the standard derivation of the kinetic equation using our above notations. The derivation of the kinetic equation used in the literature (see references in the Introduction section) involves a slow amplitude assumption and a separate assumption leading to a Gaussian decomposition of a fourth-order correlation. The two approximations are independent and can be taken in several ways. Start by defining a slow amplitude as,
\begin{linenomath}
\begin{align}
    \hat{A}_{a,\mathbf{k}}=A_{a,\mathbf{k}}e^{ia\omega_{\mathbf{k}}t}.
\end{align}
\end{linenomath}
To simplify the RHS of the energy equation \eqref{eq:energy_equation_appendix}, we need an equation for the third moment, which, after neglecting viscosity and diffusion, is
\begin{linenomath}
\begin{align}\label{eq:3rdm}
 \frac{\partial}{\partial t}\left(\hat{A}_{a,\mathbf{k}}^{\ast}\hat{A}_{b,\mathbf{p}}
 \hat{A}_{c,\mathbf{q}}\right) 
  &=e^{-i\Delta\omega(\mathbf{k},\mathbf{p},\mathbf{q},a,b,c)t}
  \epsilon\mathcal{N}_{abc}\left( \mathbf{k,p,q}\right).
\end{align}
\end{linenomath}
where $\Delta\omega(\mathbf{k},\mathbf{p},\mathbf{q},a,b,c)\equiv\left(a\omega_{\mathbf{k}}-b\omega_{\mathbf{p}}-c\omega_{\mathbf{q}}\right)$ and $\mathcal{N}_{abc}\left(\mathbf{k,p,q}\right)$ on the RHS is given by
\begin{linenomath}
\begin{align}
   \label{eq:N}
   \mathcal{N}_{abc}\left(\mathbf{k,p,q}\right)  &=\frac{1}{4\left(2\pi\right) ^{2}}
   \displaystyle{\int\limits_{-\infty}^{\infty}}
   d^{2}\mathbf{n}d^{2}\mathbf{r}\delta \left( \mathbf{k-r-n}\right)
   \mathsf{M}_{adf}^{\ast }\left( \mathbf{k,r,n}\right) 
   A_{d,\mathbf{r}}^{\ast}A_{f,\mathbf{n}}^{\ast}A_{b,\mathbf{p}}A_{c,\mathbf{q}}
   \nonumber \\
   &+\frac{1}{4\left( 2\pi \right) ^{2}}
   \displaystyle{\int\limits_{-\infty}^{\infty}}
   d^{2}\mathbf{n}d^{2}\mathbf{r}
   \delta\left(\mathbf{p-r-n}\right) \mathsf{M}_{bdf}\left(\mathbf{p,r,n}\right)
    A_{a,\mathbf{k}}^{\ast}A_{c,\mathbf{q}}A_{d,\mathbf{r}}A_{f,\mathbf{n}}
   \nonumber  \\
   &+\frac{1}{4\left( 2\pi \right) ^{2}}
   \displaystyle{\int\limits_{-\infty}^{\infty}}
   d^{2}\mathbf{n}d^{2}\mathbf{r}
   \delta\left( \mathbf{q-r-n}\right) \mathsf{M}_{cdf}\left( \mathbf{q,r,n}\right)
   A_{a,\mathbf{k}}^{\ast }A_{b,\mathbf{p}}A_{d,\mathbf{r}}A_{f,\mathbf{n}}. 
\end{align}
\end{linenomath}
At this stage, an ensemble average is taken over both sides of equation \eqref{eq:3rdm}, and one makes use of assumptions of spatial homogeneity (translation invariance), and assuming that right- and left-propagating waves do not interact to write, 
\begin{linenomath}
\begin{align}
  \label{eq:right_left_non_interacting}
  \left\langle A_{a,\mathbf{p}}A_{b,\mathbf{q}}
  \right\rangle =V\delta_{ \mathbf{p,-q}}\delta_{a,b} E_{a}\left( \mathbf{p}\right),
\end{align}
\end{linenomath}
$V$ is the volume of the system and $E_{a,\mathbf{k}}=\frac12\langle(u_\mathbf{k}^2+w_\mathbf{k}^2+b_\mathbf{k}^2)\rangle/V$ is the energy density of waves propagating in direction $a$, with wavevector $\mathbf{k}$.
The final stage in the derivation of the kinetic equation is to use the Gaussian approximation to replace the fourth-order moment with products of second-order moments. Writing the formal solution of the equation for the triple wave amplitude product, equation \eqref{eq:3rdm}, and apply an ensemble average,
\begin{linenomath}
\begin{align}
  \label{eq:3rdm_solution}
  \left.\left\langle\hat{A}_{a,\mathbf{k}}^{\ast}
  \hat{A}_{b,\mathbf{p}}
  \hat{A}_{c,\mathbf{q}}\right\rangle\right|_0^t 
  =\epsilon\int\limits_0^t
  e^{-i\Delta\omega(\mathbf{k},\mathbf{p},\mathbf{q},a,b,c)t'}
  \left\langle\mathcal{N}_{abc}\left(\mathbf{k,p,q}\right)\right\rangle dt'.
\end{align}
\end{linenomath}
Finally, use the Gaussian decomposition of the four-point correlation,
\begin{linenomath}
\begin{align}
\langle \hat{A}_{a,\mathbf{r}}\hat{A}_{b,\mathbf{n}}\hat{A}_{c,\mathbf{p}}\hat{A}_{d,\mathbf{q}} \rangle
&=\langle \hat{A}_{a,\mathbf{r}}\hat{A}_{b,\mathbf{n}}\rangle\langle \hat{A}_{c,\mathbf{p}}\hat{A}_{d,\mathbf{q}} \rangle 
\nonumber \\
&+\langle \hat{A}_{a,\mathbf{r}}\hat{A}_{c,\mathbf{p}}\rangle
\langle \hat{A}_{b,\mathbf{n}}\hat{A}_{d,\mathbf{q}}\rangle
+\langle \hat{A}_{a,\mathbf{r}}\hat{A}_{d,\mathbf{q}}\rangle\langle \hat{A}_{b,\mathbf{n}}\hat{A}_{c,\mathbf{p}} \rangle.
\label{eq:Gaussian-approximation}
\end{align}
\end{linenomath}
Together, these approximations and assumptions lead to the following simplification of the nonlinear interaction term in the energy equation,
\begin{linenomath}
\begin{align}
    \left\langle\mathcal{N}_{abc}\left(\mathbf{k,p,q}
    \right)\right\rangle
    &=\frac{1}{2\left( 2\pi \right)^{2}}\delta \left( \mathbf{k-p-q}\right)  
\Big[\mathsf{M}_{abc}\left(\mathbf{k,p,q}\right)E_{b}\left( \mathbf{p}\right)E_{c}\left( \mathbf{q}\right) \nonumber \\
 &-\mathsf{M}_{bac}\left( \mathbf{p,k,q}\right)E_{a}(\mathbf{k})E_{c}(\mathbf{q})
-\mathsf{M}_{cab}\left( \mathbf{q,k,p}\right)E_{a}(\mathbf{k})E_{b}(\mathbf{p})\Big].
\end{align}
\end{linenomath}
The derivation of the above equation uses the fact that $\mathsf{M}_{abc}\left( \mathbf{k,r,r}\right)=\mathsf{M}_{abc}\left( \mathbf{k,r,-r}\right)=0$ and the fact that $\mathsf{M}_{abc}\left( \mathbf{k,p,q}\right)$ is real and $\mathsf{M}_{abc}\left( \mathbf{k,p,q}\right)=\mathsf{M}_{acb}\left( \mathbf{k,q,p}\right)$.
Substituting the solution for the third-order correlation function, in terms of the fourth-order correlation function, into the energy equation \eqref{eq:energy_equation_appendix}, 
results in the closed kinetic equation for the energy spectrum. The equation for the energy spectrum may be written as,
\begin{linenomath}
\begin{align}
   \label{eq:F5}
   \frac{\partial}{\partial t}E_{a}(\mathbf{k},t)
   &= \frac{\epsilon^2}{4(2\pi)^4}Re\Bigg(
   \int\limits_{-\infty}^{\infty}
   d^{2}\mathbf{q}d^{2}\mathbf{p}\delta\left(  \mathbf{k-p-q}\right)
   {\displaystyle\sum\limits_{b,c}}
   \mathsf{M}_{abc}\left(  \mathbf{k,p,q}\right)
   \nonumber \\
   &\int\limits_0^t dt'
   e^{i\Delta\omega(\mathbf{k},\mathbf{p},\mathbf{q},a,b,c)(t-t')}
   \Big[\mathsf{M}_{abc}\left(\mathbf{k,p,q}\right)
   E_{b}(\mathbf{p},t') E_{c}(\mathbf{q},t')\nonumber
   \nonumber\\
   &-\mathsf{M}_{bac}\left(\mathbf{p,k,q}\right)
   E_{a}(\mathbf{k},t')E_{c}(\mathbf{q},t')
   -\mathsf{M}_{cab}\left(\mathbf{q,k,p}\right)
   E_{a}(\mathbf{k},t')E_{b}(\mathbf{p},t')\Big]
   \Bigg)\nonumber\\
   &  +2Re\left(\left\langle F_{a}^{\ast}\left(  \mathbf{k}\right)  A_{a,\mathbf{k}}\right\rangle\right).
\end{align}
\end{linenomath}
One may write the integral over time that appears on the RHS of the above energy equation as
\begin{linenomath}
\begin{align}
\int\limits_0^t dt'
e^{i\left(a\omega_{\mathbf{k}}-b\omega_{\mathbf{p}}-c\omega_{\mathbf{q}}\right)(t-t')}\mathcal{G}(\mathbf{k},\mathbf{p},\mathbf{q},a,b,c,t') dt',
\label{eq:RHS-of-3-product-using-G(E)}
\end{align}
\end{linenomath}
where the slow nonlinear interaction is represented by,
\begin{linenomath}
\begin{align}
    &\mathcal{G}(\mathbf{k},\mathbf{p},\mathbf{q},a,b,c,t')=
    \Big[\mathsf{M}_{abc}\left(\mathbf{k},\mathbf{p},\mathbf{q}\right)
 E_{b}(\mathbf{p},t') E_{c}(\mathbf{q},t')\nonumber
\nonumber\\
 &-\mathsf{M}_{bac}\left(\mathbf{p,k,q}\right)
 E_{a}(\mathbf{k},t')E_{c}(\mathbf{q},t')
-\mathsf{M}_{cab}\left(\mathbf{q,k,p}\right)
E_{a}(\mathbf{k},t')E_{b}(\mathbf{p},t')\Big].
\label{eq:define-G}
\end{align}
\end{linenomath}
The final steps of assuming that the ensemble-averaged energy spectrum varies slowly in time, taking the limit $t\to\infty$, and taking the slowly varying ensemble-averaged energies out of the time integral, lead to the resonant interaction constraint on the frequencies of interacting waves,
\begin{linenomath}
\begin{align}
&\int\limits_0^t
e^{i\left(a\omega_{\mathbf{k}}-b\omega_{\mathbf{p}}-c\omega_{\mathbf{q}}\right)(t-t')}dt'\;
\frac1t \int\limits_0^t \mathcal{G}(\langle{E(t'')}\rangle) dt''
\nonumber \\
&=\delta(a\omega_{\mathbf{k}}-b\omega_{\mathbf{p}}-c\omega_{\mathbf{q}})
(\pi/t)\int\limits_0^t \mathcal{G}(\langle{E(t'')}\rangle) dt''.
\label{eq:split-int-G}
\end{align}
\end{linenomath}
Substituting eqs.~\eqref{eq:define-G} and \eqref{eq:split-int-G} into eq.~\eqref{eq:F5}, yields the kinetic equation.

\cite{Lvov-Lvov-Newell-et-al-1997:statistical} used a similar derivation, including the effects of nonlinear broadening by making the frequencies complex, thereby introducing damping \cite[][]{Majda-McLaughlin-Tabak-1997:one} into the integral in eq.~\eqref{eq:RHS-of-3-product-using-G(E)}. As a result, the integral leads to a Lorentzian, allowing for non-resonance interactions weighted by their distance from the resonance curves rather than a delta function. The specific form of the broadening is difficult to estimate. For example, \cite{Lvov-Polzin-Yokoyama-2012:resonant} considered broadening proportional to the frequency, while \cite{Pan-Yue-2017:understanding, Zhang-Pan-2022:numerical} considered a frequency-independent broadening.

\clearpage\newpage
 
\section{Supplementary Figures and Tables}
\label{sec:appendix-supplementary-figures}

\renewcommand\thefigure{S.\arabic{figure}}    
\renewcommand\thetable{S.\arabic{table}}    
\setcounter{figure}{0}

Following are the supplementary Table and Figures referred to in the main text.

\clearpage
\begin{table} 
  \caption{\textbf{Showing the robustness of the slowness criterion $V_1$ used in the paper.} The time scale (in days) used to calculate the slowness measure $V_1$ (eq.~\ref{eq:V1}) based on different measures for the wave amplitude variability. In all cases, the time scale is evaluated as in eq.~\eqref{eq:V1} using $2\pi{[\hat{A}_{a,\mathbf{k}}]}/\left[{d\hat{A}_{a,\mathbf{k}}}/{dt}\right]$. The different table rows correspond to using $[\cdot]$ as the RMS of the real part, the imaginary part, and the absolute value of the \textit{slow} wave amplitude and of its time derivative. The last row is the linear wave time scale based on the dispersion relation.}
  \medskip
  \label{tab:time-scale-A-and-abs_A}
  \centering
  \begin{tabular}{lccc}
    \toprule
           \diagbox{$[\cdot]$}{$\mathbf{k}=$} 
                      & $(40,4)$ & $(12,24)$ & $(24,12)$ \\
    \midrule
    Real part         &   0.31        &   0.77    &    0.55        \\
    Imaginary part    &   0.34        &   0.74    &    0.60       \\
    Absolute value    &   0.51        &   1.1     &    0.85       \\
    $2\pi/\omega$     &   0.21        &   4.2    &     1          \\
   \bottomrule
  \end{tabular}
\end{table}
\clearpage

\begin{figure*}
\includegraphics[width=\textwidth]{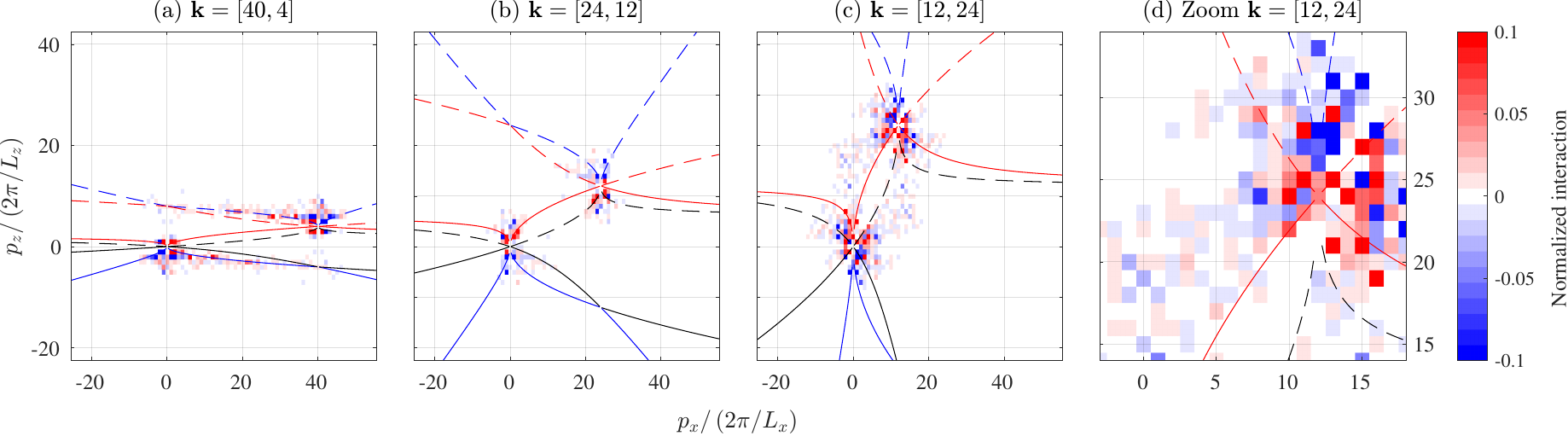}
\caption{(a--c) Time-average of the interaction term (the expression in blue on the second line of eq.~\eqref{eq:energy-equation} in the main text) for $\mathbf{k}=(40,4),~(24,12)$, and $(12,24)$, for the \textbf{weakly forced} run, normalized (for each $\mathbf{k}$) by its maximum value, \textbf{zooming in the contour range close to zero to see a wider range of interacting wavevectors}. (d) A zoom into the region of $\mathbf{p}\sim\mathbf{k}$ for panel (c). The lines represent the resonance curves.}
\label{fig:interaction-terms_28_climn=0.1}
\end{figure*}

\begin{figure*}
\includegraphics[width=\textwidth]{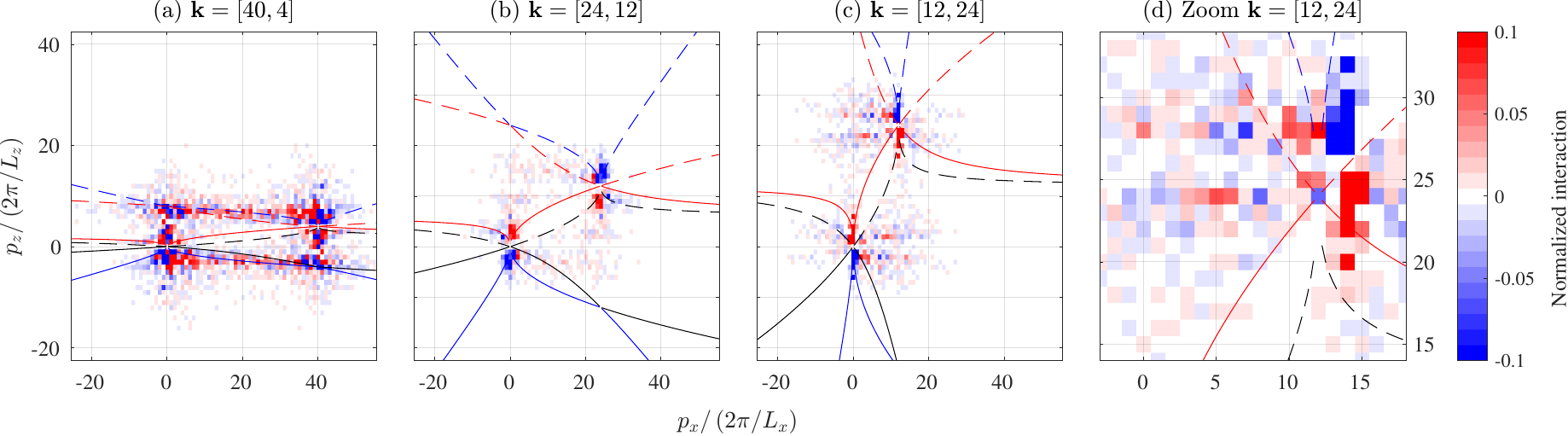}
\caption{(a--c) Time-average of the interaction term (the expression in blue on the second line of eq.~\eqref{eq:energy-equation}) 
for $\mathbf{k}=(40,4),~(24,12)$, and $(12,24)$, for the \textbf{strongly forced} run, normalized (for each $\mathbf{k}$) by its maximum value, \textbf{zooming in the contour range close to zero to see a wider range of interacting wavenumbers}. (d) A zoom into the region of $\mathbf{p}\sim\mathbf{k}$ for panel (c). The lines represent the resonance curves.}
\label{fig:interaction-terms_29_climn=0.1}
\end{figure*}

\begin{figure*}
\includegraphics[width=\textwidth]{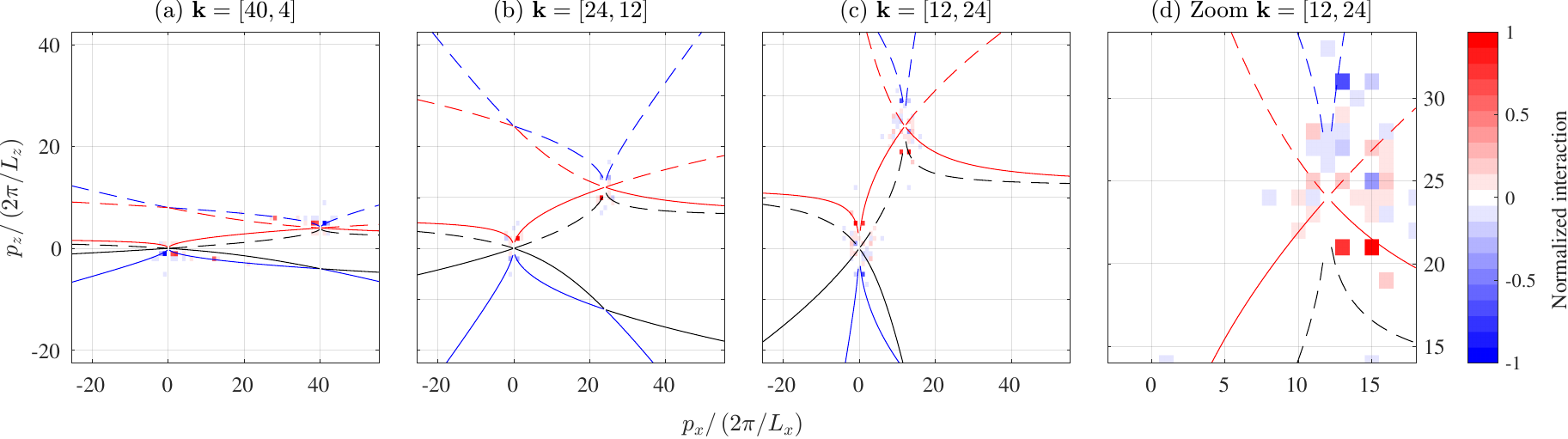}
\caption{(a--c) Time-average of the interaction term (the expression in blue on the second line of eq.~\eqref{eq:energy-equation} in the main text) for $\mathbf{k}=(40,4),~(24,12)$, and $(12,24)$, for the \textbf{weakly forced} run, normalized (for each $\mathbf{k}$) by its maximum value. (d) A zoom into the region of $\mathbf{p}\sim\mathbf{k}$ for panel (c). The lines represent the resonance curves.}
\label{fig:interaction-terms_28_climn=1}
\end{figure*}

\begin{figure*}
\includegraphics[width=\textwidth]{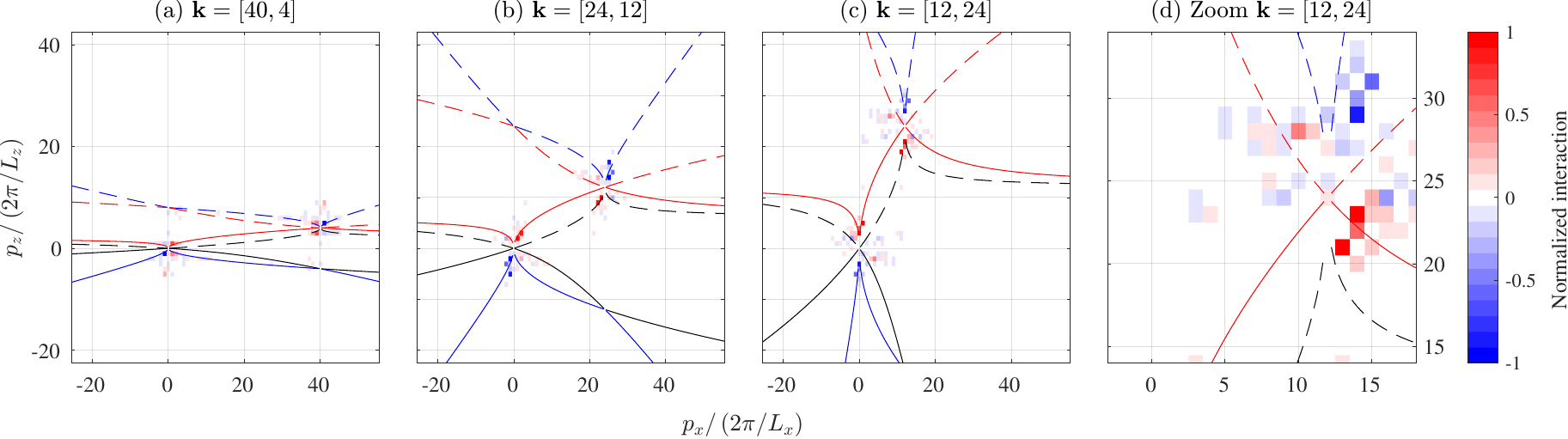}
\caption{(a--c) Time-average of the interaction term (the expression in blue on the second line of eq.~\eqref{eq:energy-equation} in the main text) for $\mathbf{k}=(40,4),~(24,12)$, and $(12,24)$, for the \textbf{medium forced} run, normalized (for each $\mathbf{k}$) by its maximum value. (d) A zoom into the region of $\mathbf{p}\sim\mathbf{k}$ for panel (c). The lines represent the resonance curves.}
\label{fig:interaction-terms_27_climn=1}
\end{figure*}

\begin{figure*}
\includegraphics[width=\textwidth]{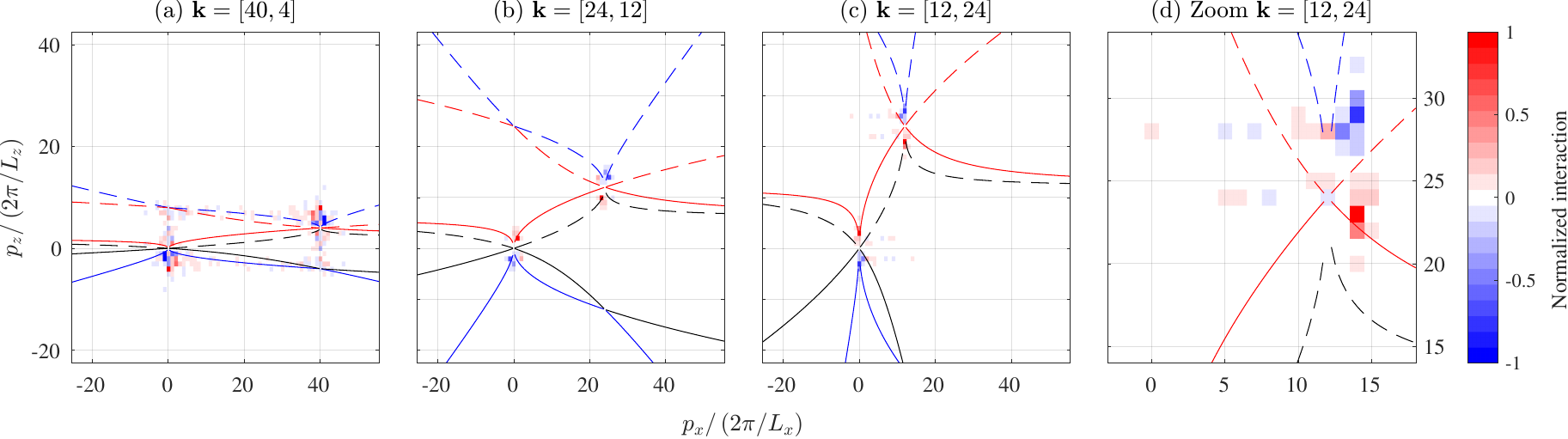}
\caption{(a--c) Time-average of the interaction term (the expression in blue on the second line of eq.~\eqref{eq:energy-equation}) 
in the main text for $\mathbf{k}=(40,4),~(24,12)$, and $(12,24)$, for the \textbf{strongly forced} run, normalized (for each $\mathbf{k}$) by its maximum value. (d) A zoom into the region of $\mathbf{p}\sim\mathbf{k}$ for panel (c). The lines represent the resonance curves.}
\label{fig:interaction-terms_29_climn=1}
\end{figure*}

\begin{figure*}
\centering
\includegraphics[width=\textwidth]{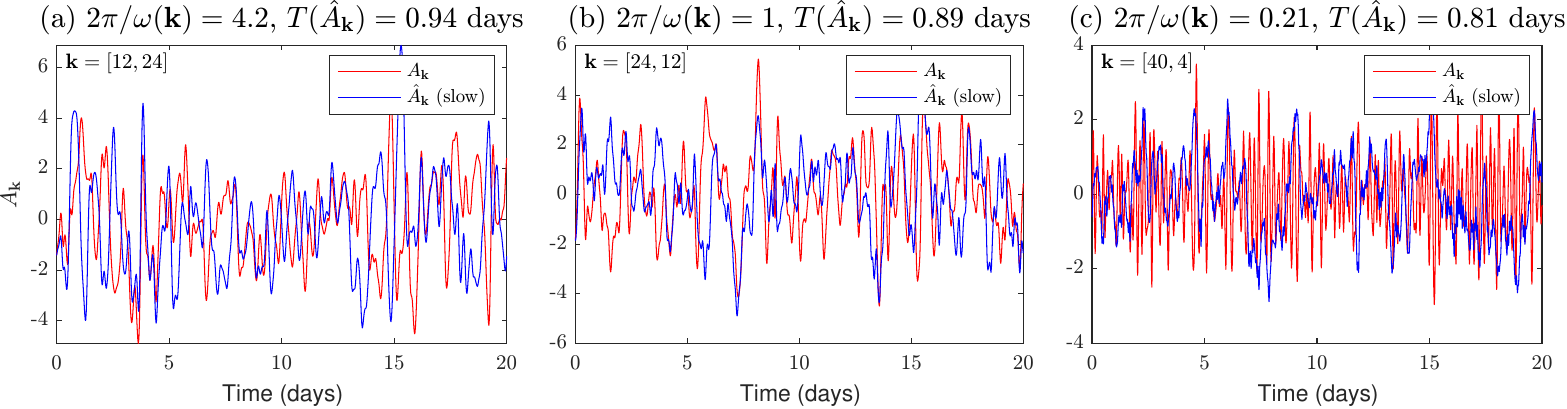}
\caption{Time series of the fast and slow amplitudes of three particular wavevectors in the \textbf{medium-amplitude forced} simulation. These panels demonstrate that the slow amplitude (blue) varies on a time scale not much longer than that of the linear wave period, as also quantified in Fig.~\ref{fig:V1}.}
\label{fig:Fig_Ap_Ap_slow_timeseries_expnum_27}
\end{figure*}

\begin{figure*}
\centering
\includegraphics[width=\textwidth]{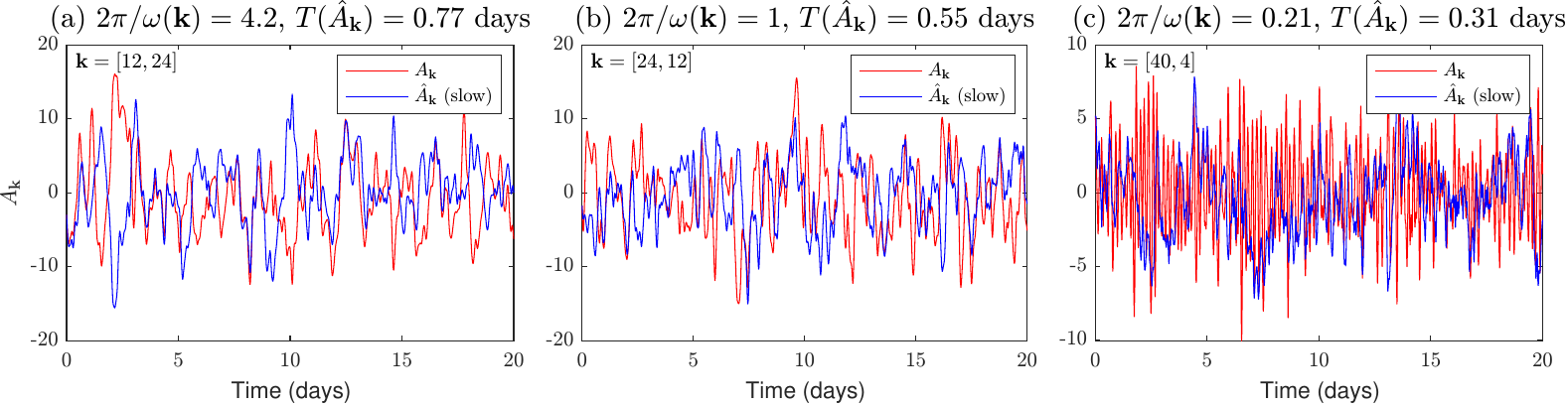}
\caption{Time series of the fast and slow amplitudes of three particular wavevectors in the \textbf{strongly forced} simulation. These panels demonstrate that the slow amplitude (blue) varies on a time scale not much longer than that of the linear wave period, as also quantified in Fig.~\ref{fig:V1}.}
\label{fig:Fig_Ap_Ap_slow_timeseries_expnum_29}
\end{figure*}

\begin{figure*}
\includegraphics[width=\textwidth]{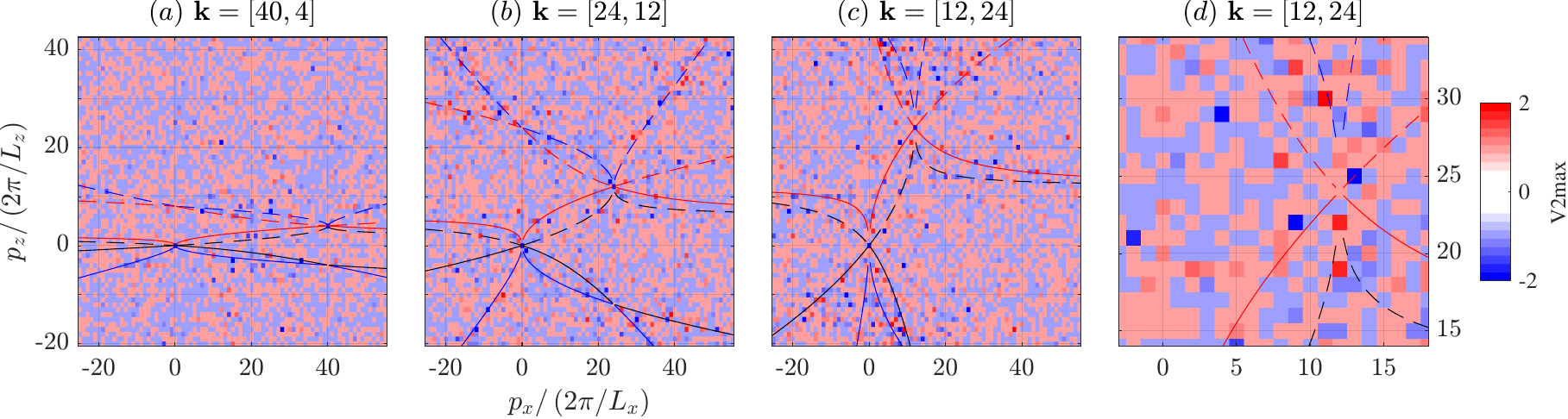}
\caption{Testing whether the three-product of wave amplitudes slowly varies (criterion $V_2$, eq.~\eqref{eq:V2} in the main text) for the \textbf{medium forcing} case, for the denoted wavevectors.}
\label{fig:V2_27}
\end{figure*}

\begin{figure*}
\includegraphics[width=\textwidth]{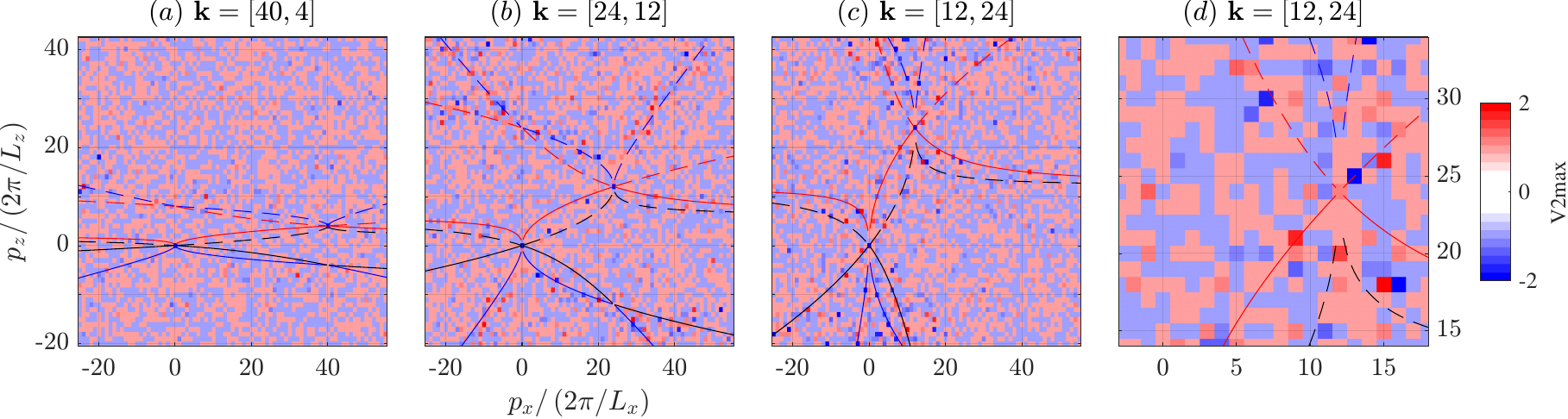}
\caption{Testing whether the three-product of wave amplitudes slowly varies (criterion $V_2$, eq.~\eqref{eq:V2} in the main text) for the \textbf{strong forcing} case, for the denoted wavevectors.}
\label{fig:V2_29}
\end{figure*}

\begin{figure*}
\includegraphics[width=\textwidth]{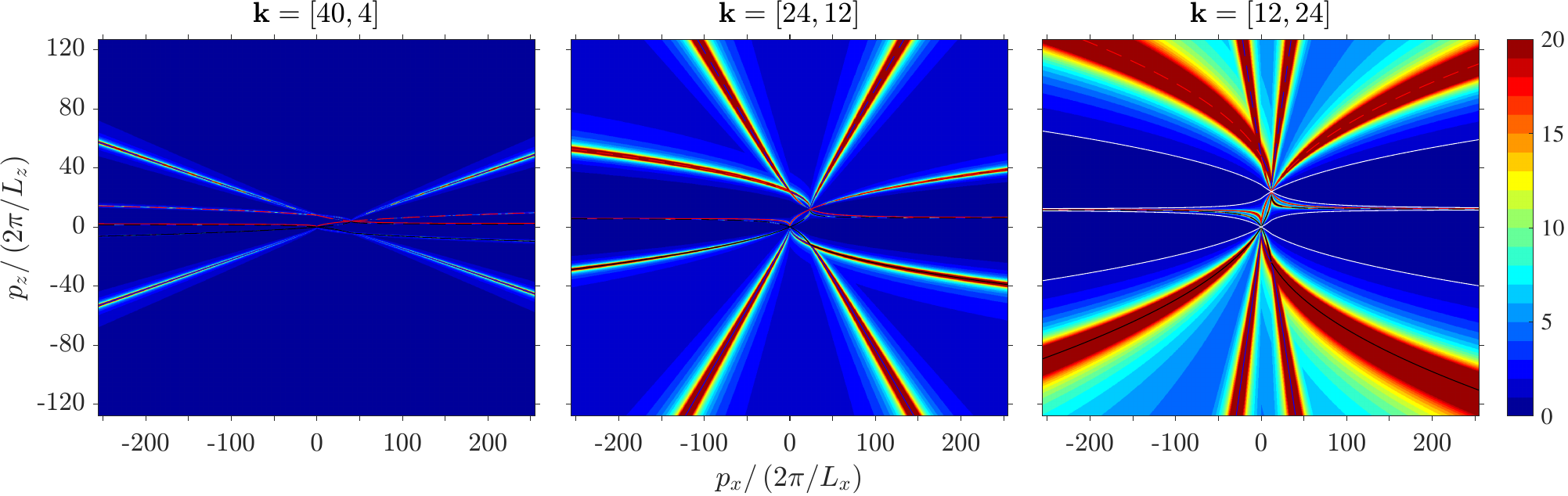}
\caption{The period of the oscillating term in the equation for the three-product of the slow wave amplitudes eq.~\eqref{eq:3rdm_solution}, $2\pi/\Delta\omega$, in units of days (criterion $V_3$, equation \eqref{eq:V3}, using $T=1$ day). On the resonant curves, this period diverges, and it is also not small in their vicinity, in seeming contradiction to the assumption used to derive the resonant condition on the frequencies of the interacting waves. The period is shown for three wavevectors, $\mathbf{k}=(40,4)$, $(24,12)$, $(12,24)$. The values shown represent the maximum of $V_3$ with respect to the indices $a,b,c$.}
\label{fig:V3}
\end{figure*}
\newpage

\begin{figure*}
\includegraphics[width=\textwidth]{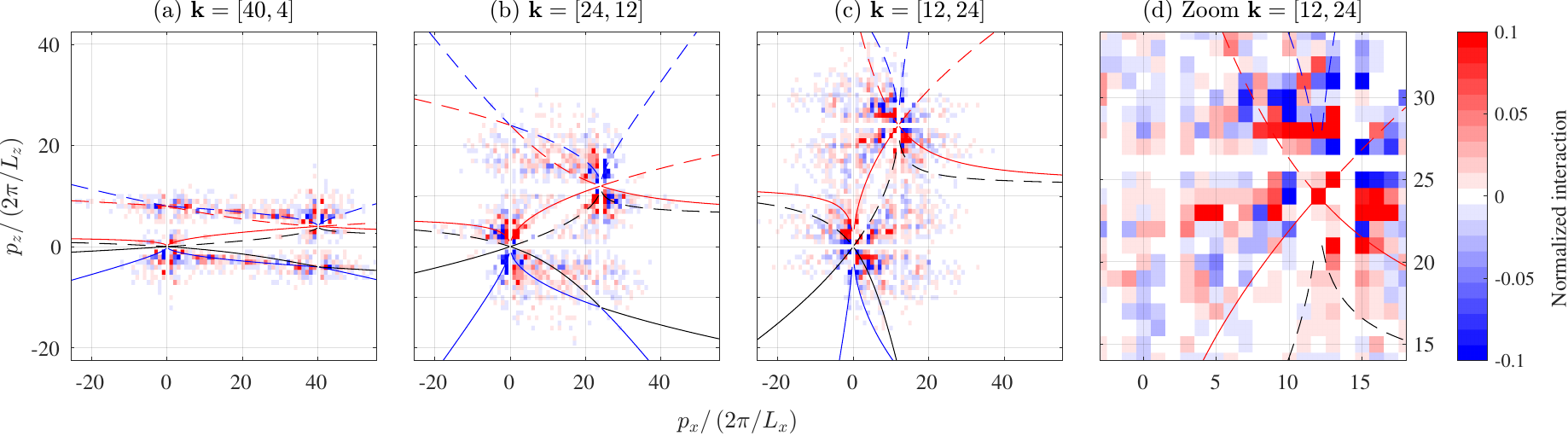}
\caption{(a--c) Time-average of the interaction term (the expression in blue on the second line of eq.~\eqref{eq:energy-equation}) 
for $\mathbf{k}=(40,4),~(24,12)$, and $(12,24)$, for the \textbf{weakly forced doubled resolution} run, with $\mathbf{u}$ damped to zero for $k_x=0$ and all $k_z$, and also damped to zero for $k_z=0$ and all $k_x$. The shown interaction values are normalized (for each $\mathbf{k}$) by their maximum, \textbf{zooming in the contour range close to zero to see a wider range of interacting wavenumbers}. (d) A zoom into the region of $\mathbf{p}\sim\mathbf{k}$ for panel (c). The lines represent the resonance curves.}
\label{fig:interaction-terms_3_climn=0.1}
\end{figure*}

\end{document}